\documentclass[aps,prd,preprintnumbers,nofootinbib,a4paper,11pt]{revtex4-2}
\usepackage[margin=1.1111023622in]{geometry}
\pdfoutput=1

\usepackage{graphicx}

\usepackage{url}
\usepackage{bm}
\usepackage{bbm}
\usepackage{caption}
\usepackage{feynmp-auto}
\usepackage{pifont}
\usepackage{ytableau}
\usepackage{array}
\usepackage{amsfonts}
\usepackage{amsmath}
\usepackage{amssymb}
\usepackage{physics}
\usepackage{colortbl}
\usepackage{diagbox}
\usepackage{float}

\usepackage[bookmarks, pagebackref=false]{hyperref}
\usepackage{color}
	\definecolor{rossoCP3}{cmyk}{0,.88,.77,.40}
		\definecolor{graa}{rgb}{0.8,0.8,0.8}
		\definecolor{blaa}{rgb}{0.2,0.2,0.6}
		\hypersetup{
			colorlinks,
			bookmarksopen,
			bookmarksnumbered,
			citecolor=blaa, 
			linkcolor=rossoCP3,	
			urlcolor=rossoCP3,	
			}
\usepackage{cleveref}
\newcommand{\drawsquare}[2]{\hbox{%
\rule{#2pt}{#1pt}\hskip-#2pt
\rule{#1pt}{#2pt}\hskip-#1pt
\rule[#1pt]{#1pt}{#2pt}}\rule[#1pt]{#2pt}{#2pt}\hskip-#2pt
\rule{#2pt}{#1pt}}

\newcommand{\fund}{\drawsquare{6.5}{0.4}}
\newcommand{\afund}{\overline{\fund}}

\newcommand{\asymm}{\raisebox{-3pt}{\drawsquare{6.5}{0.4}\hskip-6.9pt%
        \raisebox{6.5pt}{\drawsquare{6.5}{0.4}}}}

\newcommand{\beq}{\begin{eqnarray}}
\newcommand{\eeq}{\end{eqnarray}}

\newcommand{\bmp}{\noindent\begin{minipage}{16cm}}
\newcommand{\emp}{\end{minipage}\vskip 7mm} 

\def\lsim{\mathrel{\rlap{\lower4pt\hbox{\hskip1pt$\sim$}}
    \raise1pt\hbox{$<$}}}                
\def\gsim{\mathrel{\rlap{\lower4pt\hbox{\hskip1pt$\sim$}}
    \raise1pt\hbox{$>$}}}                

\begin{document}
\baselineskip=15pt
\setlength{\abovedisplayskip}{10pt}
\setlength{\belowdisplayskip}{10pt}
\setlength{\abovedisplayshortskip}{4pt}
\setlength{\belowdisplayshortskip}{4pt}
\title{\Large   Hadronic Spectrum in the Chiral Large $N_c$ Extension of Quantum Chromodynamics} 
\author{Axel Halgaard Kristensen}\email{axkri19@student.sdu.dk}
\author{Thomas A. Ryttov}\email{ryttov@cp3.sdu.dk}
\affiliation{{  \rm CP}$^{\bf 3}${ \rm-Origins}, University of Southern Denmark, Campusvej 55, 5230 Odense M, Denmark}
\pagenumbering{gobble}
\begin{abstract}
We study Quantum Chromodynamics in the chiral large $N_c$ limit which contains a left-handed Weyl fermion in the fundamental representation, a left-handed Weyl fermion in the two index antisymmetric representation and $(N_c-3)$ left-handed Weyl fermions in the antifundamental representation of the $SU(N_c)$ gauge group. We construct gauge singlet composite operators and study their masses and correlation functions at large $N_c$.

It is shown that all hadron masses scale as $\sim N_c^0 n_q$ where $n_q$ is the number of constituent quarks in the hadron.
In addition by simple gluon exchange considerations it is seen that scattering amplitudes between hadrons have the same $N_c$ scaling as the mass of the lightest hadron involved. This is the case provided the hadrons in the scattering amplitude share a sufficiently large number of constituent quarks. 

The chiral large $N_c$ extension also allows for other non-trivial processes. For instance we consider two different baryonium states that are unique to this extension and that decay via emissions of two- and three-quark hadrons. Also other non-trivial scattering processes are considered. 
Finally, we study composites made of a mix of left- and right-handed fields. We categorize multiple groups of hadrons within the full spectrum according to their flavor structure. Within these groups all $n$-point functions scale the same. 

\end{abstract}

\maketitle

\newpage

\tableofcontents

\newpage
\section{Introduction}

\pagenumbering{arabic}
The strong interactions of quarks and gluons continue to be one of the most difficult phenomena to understand even more than 50 years after the birth of Quantum Chromodynamics (QCD) \cite{Gross:AsymptoticFreedom,Politzer:AsymptoticFreedom,Gross:AsymptoticFreedom1,Gross:AsymptoticFreedom2}. Great effort has been put into understanding the spectrum and physics at low and intermediate energies using lattice simulations \cite{Wilson:lattice,Creutz:lattice}, higher order perturbation theory  which is currently known to five loops \cite{Baikov:5loop,Herzog:5loop} for the running of the gauge coupling, large $N_c$ techniques \cite{tHooft_LargeN,Witten_Baryons}, holography and light cone quantization \cite{Brodsky:lightcone1,Brodsky:lightcone2}. The literature is immense but a few recent reviews on several aspects of QCD can be found here \cite{50Years_QCD,QCDreview,Sannino:Review}.

In this work we are particularly interested in studying QCD in a large number of colors $N_c$ approach. In QCD with $N_c=3$ the quarks are in the fundamental representation of the $SU(N_c=3)$ gauge group. In the original large $N_c$ extension of QCD \cite{tHooft_LargeN,Witten_Baryons,Cohen-Lebed_Tetraquarks_in_QCDf,Manohar-Dashen-Jenkins_1/N-expansion-for-baryons,Jenkins-Lebed_Baryon-mass-splittings,Dashen-Jenkins-Manohar_Spin-flavor-structure-of-large-N-baryons} the quarks are kept in the fundamental representation of the $SU(N_c)$ gauge group. We will refer to this extension as the 't Hooft limit. At infinite $N_c$ there are a number of simplifications such as the suppression of non-planar diagrams and quark loops. On the other hand it also adds a complication since baryons contain $N_c$ quarks and therefore an infinite product of quarks at infinite $N_c$. 

Another extension is to consider the 't Hooft limit but with $N_f$ pairs of fundamental quarks and then taking the large $N_f$ limit with $N_f/N_c$ kept fixed \cite{Veneziano_Large_Nf_limit,Luty_Baryons_many-colors-and-flavors}. We will refer to this as the Veneziano limit. This limit is used in \cite{Jarvinen:Holography_VQCD1,Jarvinen:Holography_VQCD2,Jarvinen:Holography_VQCD3,Alho-Jarvinen-Kiritsis:FiniteTemperature_VQCD} in a holographic approach to QCD.

At $N_c=3$ the two index antisymmetric representation is equivalent to the antifundamental  representation. So to remedy the baryon problem just mentioned, Corrigan and Ramond proposed to add a quark in the two index antisymmetric representation in addition to the fundamental quarks \cite{CorriganRamond_Note_on_Quark_content,Kiritsis-Papavassiliou_CR_limit_revisited,Hoyos-Karch_CR-baryons-and-holography,Armoni-Shore-Shifman_CR-quark-condensate,Armoni-Patella_CR-Regge-slopes}. As one extends QCD to large $N_c$ one can then form a baryon containing only three quarks at any $N_c$. We will refer to this extension as the Corrigan-Ramond extension. If the quarks belong to the two index antisymmetric representation then the quark loops are not suppressed at infinite $N_c$ and are on an equal footing with the gluons in terms of counting of the number of degrees of freedom.

Now it is also possible to consider having only two index antisymmetric quarks and no fundamental quarks at any $N_c$ \cite{Bolognesi_Orientifold-Skyrmion, Cherman-Cohen_The_Skyrmion_strikes_back,Cherman-Cohen-Lebed_All-you-need-is-N,Cherman-Cohen_strangequark-a-tale-of-two-skyrmions}. At $N_c=3$ this is still ordinary QCD. We will refer to this extension of QCD as the orientifold limit. If we consider a single quark in the two index antisymmetric representation at infinite $N_c$ this extension shares parts of its bosonic sector with $\mathcal{N}=1$ supersymmetric Yang-Mills theory \cite{Armoni-Shifman-Veneziano_Exact_results_Orientifold,Veneziano-Armoni-Shifman_Orientifold2,Veneziano-Armoni-Shifman_Orientifold3,Veneziano-Armoni-Shifman_Orientifold4,Veneziano-Armoni-Shifman_Orientifold5}.

In this work we consider a large $N_c$ extension of QCD first proposed in \cite{Ryttov_HiddenQCD} that in a certain sense stands in between the 't Hooft and orientifold large $N_c$ limits and simultaneously share certain features with the Veneziano limit. It does so in a non-trivial way since it is a chiral extension of $SU(N_c=3)$ QCD and one has to be careful about possible gauge anomalies. It is in this little honey hole between the above mentioned time honored large $N_c$ extensions our investigations will unfold. Our results are complementary to the results derived in the past for the various large $N_c$ limits.

The work is organized as follows. In section \ref{Chiral} we introduce the large $N_c$ chiral extension of QCD, in section \ref{Observations} we comment on a number of different properties of the chiral extension and in section \ref{counting} we give a brief review of large $N_c$ counting techniques. In section \ref{hadron} we provide a detailed analysis of the composite spectrum, their masses and various interactions. We finally end with the conclusions in \ref{conclusions}.

\section{The Chiral Large $N_c$ Extension}\label{Chiral}

\begin{figure}[H]
\centering
  \includegraphics[width=.4\linewidth]{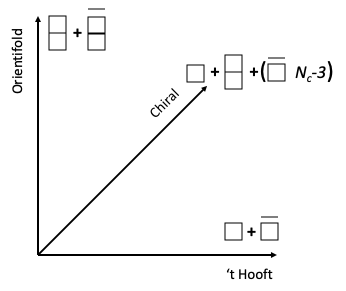}
  \caption{\textit{The three different large $N_c$ limits of QCD}}
  \label{fig}
\end{figure}

The chiral large $N_c$ extension is a third limit of QCD which in some sense is in between the 't Hooft and orientifold limits. Schematically it can be seen in Fig. \ref{fig} which is taken from \cite{Ryttov_HiddenQCD}. In the chiral large $N_c$ limit we consider a gauge theory with gauge group $SU(N_c)$ and with associated gauge fields $\left( A_{\mu} \right)^{i}_{\phantom{i}j}$ in the adjoint representation where $\mu=0,\ldots,3$ is a Lorentz index and $i,j=1,\ldots,N_c$ are $SU(N_c)$ gauge indices. 
We also add a single left-handed Weyl fermion $q_{\alpha}^{i}$ in the fundamental representation of the gauge group and a single left-handed Weyl fermion $\tilde{q}_{\alpha}^{[ij]} = -\tilde{q}_{\alpha}^{[ji]}$ in the two index antisymmetric representation of the gauge group. 
This is a chiral gauge theory, and in order to cancel the gauge anomaly we add $(N_c-3)$ left-handed Weyl fermions $Q_{\alpha,i,f}$ in the antifundamental representation of the gauge group. Here the index $\alpha$ is an $SL(2,\mathbb{C})$ spinor index and $f$ is a global $SU(N_c-3)$ index. We will refer to $f$ as a flavor index. The theory is a generalized Georgi-Glashow model \cite{Georgi-Glashow_gGG_models} and is chiral for $N_c>3$. We here summarize the particle content

\begin{table}[H]
\centering
\begin{tabular}{ |c|c|c|c|c| } 
\hline
  & $\left[ SU(N_c) \right] $ & $SU(N_c-3) $ & $U_1(1)$ & $U_2(1)$ \\ 
  \hline
 $q_{\alpha}$ & $\fund$ & $1$ & $N_c-2$ & $-(N_c-1)$   \\ 
 $\tilde{q}_{\alpha}$ & $\asymm$ & 1 & $N_c-4$ & 2 \\ 
 $Q_{\alpha}$ & $\afund$ & $\afund$ & $-(N_c-2)$ & -1 \\
 $A_{\mu}$  & \text{adj} & 1 & 0 & 0  \\ 
 \hline
\end{tabular}
\caption{\textit{This table summarizes the particle content of the chiral extension, and shows the charges of the two anomaly free $U(1)$ symmetries.}}
\label{tab:content}
\end{table}
The two abelian $U(1)$ symmetries are both anomaly free\footnote{The contribution from $n$ left-handed Weyl fermions to a $U(1)$ anomaly is $T_r Q n $ where $T_r$  is the trace normalization factor of the gauge group representation $r$ (see appendix \ref{app:group}) and $Q$ is the $U(1)$ charge.}
\begin{align}
& U_1(1): \qquad \frac{1}{2}(N_c-2)+\frac{N_c-2}{2}2 - \frac{1}{2} (N_c-2) (N_c-3) = 0 \label{eq:U1symmetries_1}\\
& U_2(1): \qquad -\frac{1}{2} (N_c-1) + \frac{N_c-2}{2}  2 - \frac{1}{2} (N_c-3) = 0 \label{eq:U1symmetries_2}
\end{align}
Most importantly for $N_c=3$ the Weyl fermions $Q_{\alpha,i,f}$ disappear from the spectrum and the two index antisymmetric representation is equivalent to the antifundamental representation. Therefore the theory is vector-like and is $SU(N_c=3)$ QCD with a single massless Dirac quark flavor
\begin{align}
\Psi_D = 
\begin{pmatrix}
q_{\alpha}^{i} \\
\epsilon^{ijk} \epsilon^{\dot{\alpha} \dot{\beta} } \bar{\tilde{q}}_{\dot{\beta}, [jk]}
\end{pmatrix}
\end{align}
Our convention for Hermitian conjugation is 
\begin{align}
\left( \tilde{q}_{\alpha}^{[ij]} \right)^{\dagger} = \bar{\tilde{q}}_{\dot{\alpha},[ij]}
\end{align}

The model above provides a third alternative large $N_c$ limit for one flavor QCD. First note that it partly resembles the 't Hooft limit since it contains a Weyl fermion $q_{\alpha}^{i}$ in the fundamental representation of the gauge group. Second note that it partly resembles the orientifold limit since it contains a Weyl fermion $\tilde{q}_{\alpha}^{[ij]}$ in the two index antisymmetric representation of the gauge group. In this sense the large $N_c$ limit we are considering stands in between these two time honored limits. Now third note that to non-trivially cancel the gauge anomalies we add $N_c-3$ Weyl fermions in the antifundamental representation. Taking the large $N_c$ limit of these fermions then in some sense resembles the Veneziano limit where one takes the limit of a large number of flavors. As a final remark please note that we can trivially extend the large $N_c$ limit of one flavor QCD to that of $N_f$ flavor QCD by just adding $N_f$ copies of the same fermion content. 

Various possible candidate phases for generalized Georgi-Glashow models have been studied in \cite{Appelquist:Phases}. In general the phase structure of chiral gauge theories has been of significant interest in the past. See for instance \cite{Ryttov_HiddenQCD,Appelquist:PhasesShrock1,Appelquist:PhasesShrock2,Armoni:Phases,Shi:Phases,Shi:Phases2,Shi:Phases3,Bolognesi:Phases,Bolognesi:Phases2,Bolognesi:Phases3,Bolognesi:Phases4,Smith:Phases,Karasik:Phases,Anber:Phases}.

When taking the large $N_c$ limit we must be careful and make sure that we correctly normalize the fields and properly account for various $SU(N_c)$ color factors. We want to use the double line notation and hence we consider the gauge fields and the Weyl fermions in the two index antisymmetric representation as rank-2 tensor representations. We also want to rescale the fields so that the gauge coupling only appears as an overall factor in the Lagrangian. We discuss this in detail in appendix \ref{app:group} and \ref{app:canonicalL}. With these conventions the Lagrangian is
\begin{align}
\mathcal{L}  =& \frac{N_c}{\lambda} \left[ - \frac{1}{2} \text{Tr}\ F_{\mu\nu}F^{\mu\nu} + i \bar{q} \bar{\sigma}^{\mu} D_{\mu} q 
+ 2i \text{Tr} \ \left( \bar{\tilde{q}} \bar{\sigma}^{\mu} D_{\mu} \tilde{q} \right)  
+ i \bar{Q} \bar{\sigma}^{\mu} D_{\mu} Q  \right. \nonumber \\
& \left. - \frac{1}{\xi} \text{Tr} \ \left( \partial^{\mu} A_{\mu} \right)^2 
+ 2 \text{Tr} \ \bar{c} \left( - \partial^{\mu} D_{\mu} c \right)   \right]
\end{align}
with 
\begin{align}
F_{\mu\nu} &=  F_{\mu\nu}^{a} T^a_F= \partial_{\mu} A_{\nu} - \partial_{\nu} A_{\mu} - i \left[ A_{\mu},A_{\nu} \right] \\
D_{\mu} q^i & = \partial_{\mu} q^i -i \left( A_{\mu} \right)^{i}_{\phantom{i}j} q^j \ , \qquad \left( A_{\mu} \right)^{i}_{\phantom{i}j} = A_{\mu}^{a} \left( T^a_F \right)^{i}_{\phantom{i}j} \\
\left( D_{\mu} \tilde{q}  \right)^{ij}&= \partial_{\mu} \tilde{q}^{ij} -i \left( \left( A_{\mu} \right)^i_{\phantom{i}k} \tilde{q}^{kj} + \tilde{q}^{ik}\left( A_{\mu}^T \right)_k^{\phantom{k}j}   \right) \ , \qquad \left( A_{\mu} \right)^{i}_{\phantom{i}j} = A_{\mu}^{a} \left( T^a_F \right)^{i}_{\phantom{i}j} \\
D_{\mu} Q_i & = \partial_{\mu} Q_i -i \left( A_{\mu} \right)_{i}^{\phantom{i}j} \tilde{q}_j \ , \qquad  \left( A_{\mu} \right)_{i}^{\phantom{i}j} = A_{\mu}^{a} \left( T^{a}_{\bar{F}} \right)_{i}^{\phantom{i}j} =  - A_{\mu}^{a} \left( T^{a,T}_{F} \right)_{i}^{\phantom{i}j}  \\
\left( D_{\mu} c \right)^{i}_{\phantom{i}j} &= \partial_{\mu} c^{i}_{\phantom{i}j} - i \left[ A_{\mu} , c\right]^{i}_{\phantom{i}j} \ , \qquad \left( A_{\mu} \right)^{i}_{\phantom{i}j} = A_{\mu}^{a} \left( T^a_F \right)^{i}_{\phantom{i}j} \ , \qquad \left( c \right)^{i}_{\phantom{i}j} = c^{a} \left( T^a_F \right)^{i}_{\phantom{i}j}
\end{align}
and the 't Hooft coupling being $\lambda = g^2 N_c$. We have chosen linear covariant gauge with gauge parameter $\xi$ while $c = c^a T^a_F$ are ghost fields in the adjoint representation. This Lagrangian is equivalent to the canonically normalized Lagrangian in Eq. \ref{eq:canonicalL} when appropriately rescaling the fields and carrying out the traces. 

We can obtain the propagators by multiplying the canonically normalized propagators in Eq. \ref{eq:canonicalP} by the completeness relations in Eq. \ref{eq:relations1}-\ref{eq:relations2}. We then obtain
\begin{align}\label{eq:canonicalP_2} 
\left\langle \left( A_{\mu}(x) \right)^{i}_{\phantom{j}j}  \left( A_{\nu}(y) \right)^{k}_{\phantom{k}l} \right\rangle & = \frac{1}{2}  \Delta_{\mu\nu} (x-y) \frac{\lambda}{N_c} \left( \delta^{i}_{\phantom{i}l} \delta^{k}_{\phantom{k}j} - \frac{1}{N_c} \delta^{i}_{\phantom{i}j} \delta^{k}_{\phantom{k}l}  \right) \nonumber \\
\left\langle q^{i}_{\alpha}(x) \bar{q}_{\dot{\alpha},j} (y) \right\rangle &= S_{\alpha \dot{\alpha}}(x-y) \frac{\lambda}{N_c} \delta^{i}_{\phantom{i}j} \nonumber \\ 
\left\langle \tilde{q}_{\alpha}^{[ij]}(x) \bar{\tilde{q}}_{\dot{\alpha},[kl]} (y) \right\rangle &= \frac{1}{4}S_{\alpha \dot{\alpha}}(x-y) \frac{\lambda}{N_c}  \left( \delta^{i}_{\phantom{i}l} \delta^{j}_{\phantom{j}k} -  \delta^{i}_{\phantom{i}k} \delta^{j}_{\phantom{j}l}  \right) \\
\left\langle Q_{\alpha,i,f}(x) \bar{Q}_{\dot{\alpha}}^{j,f'} (y) \right\rangle &= S_{\alpha \dot{\alpha}}(x-y) \frac{\lambda}{N_c} \delta_{i}^{\phantom{i}j} \delta_{f}^{\phantom{f}f'}  \nonumber \\
\left\langle \left( c (x) \right)^{i}_{\phantom{i}j} \left( \bar{c} (y) \right)^{k}_{\phantom{k}l} \right\rangle &= \frac{1}{2}  \Delta (x-y) \frac{\lambda}{N_c} \left( \delta^{i}_{\phantom{i}l} \delta^{k}_{\phantom{k}j} - \frac{1}{N_c} \delta^{i}_{\phantom{i}j} \delta^{k}_{\phantom{k}l}  \right) \nonumber
\end{align}
for the propagators. Neglecting the Lorentz part of the propagators and only writing the color structure we present them as the following Feynman diagrams
\begin{align*}
    \includegraphics[scale=0.5]{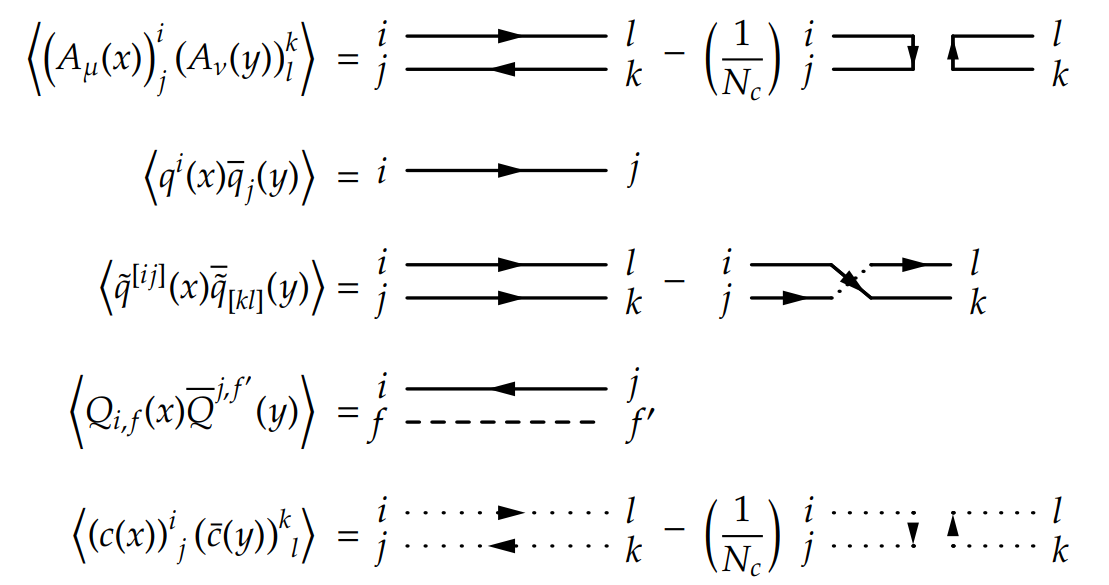}
\end{align*}
For simplicity we have not written an arrow on the flavor line for the $Q$ propagator since it will always point in the same direction as the arrow on the associated color line.

\section{Some Observations}\label{Observations}

In this section we comment on a number of observations and properties of the chiral large $N_c$ limit. First, counting of the degrees of freedom (neglecting spin) gives
\begin{align} \label{eq:DegreesOfFreedom}
\begin{matrix}
A_{\mu}: & N_c^2-1 & \xrightarrow[]{\text{large}\ N_c} & N_c^2 \\
q_{\alpha}: & N_c & \xrightarrow[]{\text{large}\ N_c} & N_c \\
\tilde{q}_{\alpha}: & \frac{1}{2}N_c(N_c-1) & \xrightarrow[]{\text{large}\ N_c} &  \frac{1}{2} N_c^2 \\
Q_{\alpha,f}: & N_c(N_c-3) & \xrightarrow[]{\text{large}\ N_c} & N_c^2 
\end{matrix}
\end{align}
So the number of degrees of freedom of the $q_{\alpha}$ Weyl fermion only grows linearly as opposed to the degrees of freedom of all the other fields $A_{\mu}, \tilde{q}_{\alpha}, Q_{\alpha,f}$ which grows quadratically. In other words the fundamental quark is suppressed.

Consider now the running of the 't Hooft coupling $\lambda = g^2 N_c$ which can be found from the running of the gauge coupling $g$. To one loop order the running is 
\begin{align}
\mu \frac{d \lambda}{d \mu} &= - 2 \lambda \left[  \tilde{b}_1 \frac{\lambda}{(4\pi)^2}  + \ldots\right] \\
3\tilde{b}_1 &= \underbrace{11}_{A_{\mu}}  - \underbrace{\frac{1}{N_c}}_{q_{\alpha}} - \underbrace{\frac{N_c-2}{N_c}}_{\tilde{q}_{\alpha}} - \underbrace{\frac{N_c-3}{N_c}}_{Q_{\alpha,f}} \qquad \xrightarrow[]{\text{large}\ N_c} \qquad  \underbrace{11}_{A_{\mu}}   - \underbrace{1}_{\tilde{q}_{\alpha}} - \underbrace{1}_{Q_{\alpha,f}}
\end{align}
So again we see that at the one loop level the fundamental quark $q_{\alpha}$ loops are suppressed by $1/N_c$ compared to the other $A_{\mu}, \tilde{q}_{\alpha}, Q_{\alpha,f}$ loops. The theory is asymptotically free at any $N_c$. 

At last consider the cancellation of the qubic gauge anomaly. The chiral large $N_c$ extension is a chiral gauge theory and the qubic gauge anomaly is
\begin{align}
SU(N_c)^3 :   \underbrace{1}_{q_{\alpha}} + \underbrace{(N_c-4)}_{\tilde{q}_{\alpha}} - \underbrace{(N_c-3)}_{Q_{\alpha,f}}  = 0 \qquad \xrightarrow[]{\text{large}\ N_c} \qquad  \underbrace{N_c}_{\tilde{q}_{\alpha}} - \underbrace{N_c}_{Q_{\alpha,f}} =0
\end{align}
Again we see that the relevance of the fundamental quark $q_{\alpha}$ is suppressed relative to $\tilde{q}_{\alpha}$ and $Q_{\alpha,f}$ in canceling the gauge anomaly.\\[-8pt]

\section{Large $N_c$ Counting}\label{counting}
The $N_c \xrightarrow[]{} \infty$ limit greatly simplifies the task of evaluating Feynman diagrams. Our most important task will be to keep track of powers of $N_c$.

In the large $N_c$ limit we take the 't Hooft coupling $\lambda$ to be fixed. As summarized in Eq. \eqref{eq:canonicalP_2} propagators come with a factor of $\frac{1}{N_c}$. Interaction vertices come with a factor of $N_c$. Each color-, and flavor-contraction comes with a factor of $\sim N_c$
\begin{align}
    \delta^i_{\phantom{i}i} = N_c, \qquad \delta^f_{\phantom{f}f} = N_c-3 \sim N_c
\end{align}

\subsection{Gluons}
We will first demonstrate how large $N_c$ counting works for a pure Yang-Mills theory. Consider a simple vacuum bubble in double-line notation
\begin{align*}
    \includegraphics{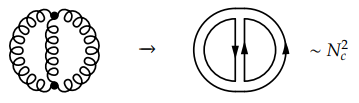}
\end{align*}
\noindent In this work we take time to flow in the upwards direction in all Feynman diagrams. To find the $N_c$ scaling of this diagram, we only need to count propagators, vertices and color-loops. There are three propagators, two vertices, and three color-loops. In total, the diagram scales like $(\frac{1}{N_c})^3\times N_c^2 \times N_c^3 \sim N_c^2$.
As examples the following diagrams are also all $\sim N_c^2$
\begin{align*}
    \includegraphics{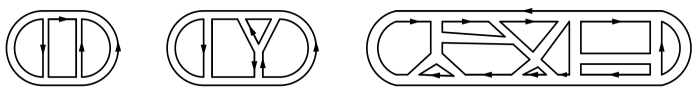}
\end{align*}
If we change the index-structure of the very first diagram, we can draw something subtly different\\[-8pt]
\begin{align*}
    \includegraphics{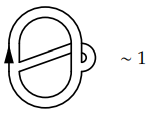}
\end{align*}
\noindent This diagram has three propagators, two vertices and \textit{one} loop, so it scales like $\sim (\frac{1}{N_c})^3 \times N_c^2 \times N_c \sim 1$. Relative to the other diagrams, this one is suppressed by $1/N_c^2$ which makes it irrelevant in the large $N_c$ limit.
The above examples hint at a hierarchy of Feynman diagrams in the large $N_c$ limit - diagrams that can be drawn on a flat plane (planar diagrams) are dominant compared to the diagrams that require a third dimension to be drawn (nonplanar diagrams). One can now construct a $1/N_c$ expansion such that the sum over all $N_c$ leading diagrams (the planar diagrams) samples a subsection of diagrams at all loop-orders. This is of course well known. \\

\subsection{Fundamental quarks}
Next, we can add $q^i$-quark lines to the vacuum bubbles. 
The counting rules are the same\\[-7pt]
\begin{align*}
    \includegraphics{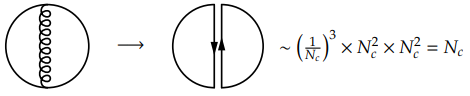}
\end{align*}
These kinds of diagrams scale like $\sim N_c$ because the quark line is a single color-line (as opposed to the gluon double-line).
If we keep the diagram planar, but add some number of internal $q^i$-quark loops
\begin{align*}
    \includegraphics{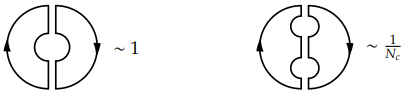}
\end{align*}
we see that each internal quark loop gives a factor of $\frac{1}{N_c}$. So internal fundamental $q^i$-quark loops are suppressed. Yet another suppressed diagram with an internal $q^i$-quark loop is\\[-8pt] 
\begin{align*}
    \includegraphics{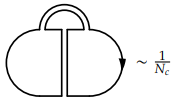}
\end{align*}
Of course, non-planar diagrams are also still suppressed.

To sum up: If we want quarks in the diagram, the planar diagrams with a single quark loop on the boundary dominate. These diagrams scale as $\sim N_c$, which is subleading compared to the gluon vacuum bubbles $\sim N_c^2$.

\subsection{Many antifundamental quarks}
The antifundamental quarks $Q_{i,f}$ have a flavor-line in addition to their color-line. The consequence is that $Q$-quark loops are not suppressed, and internal $Q$-quark loops do not alter the $N_c$ scaling
\begin{align*}
    \includegraphics{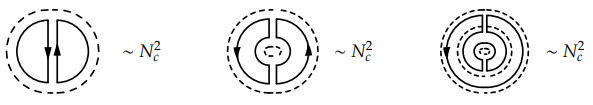}
\end{align*}
In the third diagram we see that crossing flavor-lines with fermion- or gluon-lines is not suppressed. Second, planar diagrams still dominate over non-planar ones\\[-10pt]
\begin{align*}
    \includegraphics{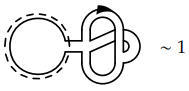}
\end{align*}
Third, interrupted $Q$-quark loops, internal or not, are suppressed. 
The reason for this rule is that the interrupted quark loops reduce the number of color-loops. This becomes obvious when we use double-line notation
\begin{align*}
    \includegraphics{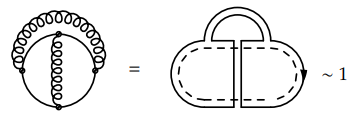}
\end{align*}

\subsection{Two-index antisymmetric quarks}
The two-index antisymmetric quarks $\tilde{q}^{[ij]}$, similarly to the gluons, have two color-lines. In contrast to the $Q$-quark loops, interrupted $\tilde{q}$-quark loops have the same scaling as uninterrupted ones. In other words, \textit{all} planar diagrams comprised of $\tilde{q}$'s and gluons, are $\sim N_c^2$ - notice that this is the same scaling as in pure Yang-Mills\\[-10pt]
\begin{align*}
    \includegraphics{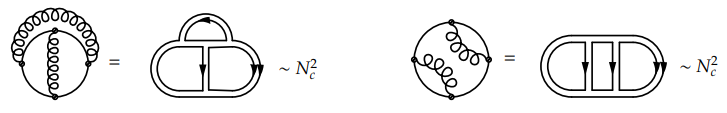}
\end{align*}
The two index antisymmetric representation is the one used in the orientifold large $N_c$ extension \cite{Armoni-Shifman-Veneziano_Exact_results_Orientifold,Veneziano-Armoni-Shifman_Orientifold2,Veneziano-Armoni-Shifman_Orientifold3,Veneziano-Armoni-Shifman_Orientifold4,Veneziano-Armoni-Shifman_Orientifold5}.\\

The vacuum bubble large $N_c$ scaling of the various quark types is in agreement with the observations discussed in section \ref{Observations}. We know how to find the $N_c$ scaling of any given diagram in the chiral large $N_c$ extension and will now move on. \\[-8pt]

\section{Hadron Spectrum}\label{hadron}
\noindent What are the minimal colorless hadrons in this theory? 
First, we have the order $N_c$-quark composites that only use one quark-type each
\begin{align}
    &B = \epsilon_{i_1...i_{N_c}} q^{i_1} \;...\; q^{i_{N_c}}, \\
    &\tilde{B}_{N_c/2} = \epsilon_{i_1...i_{N_c}}  
    \tilde{q}^{[i_1 i_2]} \;...\; \tilde{q}^{[i_{N_c-1} i_{N_c}]},    && N_c \quad \text{even}\\
    &\tilde{B}_{even} = \epsilon_{i_1...i_{N_c}} \epsilon_{j_1...j_{N_c}} 
    \tilde{q}^{[i_1 j_1]} \;...\; \tilde{q}^{[i_{N_c} j_{N_c}]},        && N_c \quad \text{even}\\
    &\tilde{B}_{odd} = \epsilon_{i_1...i_{N_c}} \epsilon_{j_1...j_{N_c}} 
    \tilde{q}^{[i_1 j_1]} \;...\; \tilde{q}^{[i_{N_c} j_{N_c}]} = 0,    && N_c \quad \text{odd}\\
    &\mathcal{B}_{f_1...f_{N_c}} = \epsilon^{i_1...i_{N_c}} Q_{i_1,f_1} \;...\; Q_{i_{N_c},f_{N_c}}
\end{align}
We have suppressed the spinor-indices. It is clear that when $N_c=3$ the composite $B = \epsilon_{ijk} q^i q^j q^k$ becomes the usual baryon $\Delta^{++}$ in one-flavor QCD. Note that the $\Tilde{B}_{odd}$ composite vanishes
\begin{align*}
    \tilde{B}_{odd} &= \epsilon_{i_1...i_{N_c}} \epsilon_{j_1...j_{N_c}} 
    \tilde{q}^{[i_1 j_1]} \;...\; \tilde{q}^{[i_{N_c} j_{N_c}]}\\
    &= (-1)^{N_c} \epsilon_{i_1...i_{N_c}} \epsilon_{j_1...j_{N_c}} 
    \tilde{q}^{[j_1 i_1]} \;...\; \tilde{q}^{[j_{N_c} i_{N_c}]}\\
    &= - \epsilon_{j_1...j_{N_c}} \epsilon_{i_1...i_{N_c}} 
    \tilde{q}^{[i_1 j_1]} \;...\; \tilde{q}^{[i_{N_c} j_{N_c}]} = -\tilde{B}_{odd} = 0
\end{align*}
We can also build order $N_c$-quark composites by combining different quark-types. One combination is
\begin{align}
    M_{(1,N_c-2)} = \epsilon_{i_1 \ldots i_{N_c}} \tilde{q}^{[i_1 i_2]} q^{i_3} \ldots q^{i_{N_c}}
\end{align}
This becomes the usual meson $M = \epsilon_{ijk} \tilde{q}^{[jk]} q^i$ when $N_c=3$ in one-flavor QCD. For general $N_c$ it is just one member of a whole family of composites. These consist of $\tilde{s}$ quarks and $s$ fundamental quarks 
\begin{align}
    M_{(\tilde{s},s)} = \epsilon_{i_1 \ldots i_{N_c}} \tilde{q}^{[i_1 i_2]} \ldots \tilde{q}^{[i_{2\tilde{s}-1} i_{2\tilde{s}}]}
    q^{i_{2s+1}} \ldots q^{i_{N_c}}, \qquad  2\tilde{s} + s = N_c 
\end{align}
As we vary $\tilde{s}$ and $s$ with $2\tilde{s} + s = N_c$ fixed, the $M_{(\tilde{s},s)}$ family of composites interpolates between the $B$ and $\tilde{B}_{N_c/2}$ composites. When $\tilde{s} = 0$ and $s = N_c$ we have $M_{(0,N_c)} = B$ and when $\tilde{s} = N_c/2$ and $s=0$ we have $M_{(N_c/2,0)} = \tilde{B}_{N_c/2}$. There are also the following two composites which consist of a fixed number of quarks not scaling with $N_c$
\begin{align}
    &X_f = Q_{i,f} q^i\\
    &Y_{ff'} = \tilde{q}^{[ij]} Q_{i,f} Q_{j,f'}
\end{align}
The $X_f$'s have the same color index contraction as the mesons in the 't Hooft limit, the difference being that they in addition have a single $SU(N_c-3)$ flavor index. The $Y_{ff'}$'s are also somewhat similar to the baryons in the Corrigan-Ramond extension \cite{CorriganRamond_Note_on_Quark_content} except for the two $SU(N_c-3)$ flavor-indices. 

Finally the Skyrmion which is denoted by $\tilde{B}$ consists of $N_c(N_c-1)/2$ quarks $\tilde{q}^{[ij]}$. They are non-trivial to write down for arbitrary $N_c$.
In $SU(3)$, the two-index antisymmetric representaion is the same as the antifundamental $\Tilde{q}_i = \epsilon_{ijk}\tilde{q}^{[jk]}$, and the baryon is
\begin{align}
    \tilde{B}_{[SU(3)]} 
    = \epsilon^{ijk} \Tilde{q}_i \Tilde{q}_j \Tilde{q}_k 
    = 2  \epsilon_{i_1j_1j_2} \epsilon_{i_2 k_1k_2}  \tilde{q}^{[i_1i_2]} \tilde{q}^{[j_1j_2]} \tilde{q}^{[k_1k_2]}
\end{align}
For $SU(4)$, 
\begin{align}
    \tilde{B}_{[SU(4)]} 
    = \bigg( &\sum_{\sigma \in S} \mbox{sign}(\sigma) 
    \epsilon_{\alpha_{\sigma(4)} \beta_{\sigma(4)} \alpha_{\sigma(2)} \alpha_{\sigma(1)}}
    \epsilon_{\alpha_{\sigma(2)} \beta_{\sigma(5)} \beta_{\sigma(2)} \alpha_{\sigma(3)}}
    \epsilon_{\alpha_{\sigma(6)} \beta_{\sigma(6)} \beta_{\sigma(1)} \beta_{\sigma(3)}} \bigg) \nonumber \\
    &\times \tilde{q}^{[\alpha_1 \beta_1]} \tilde{q}^{[\alpha_2 \beta_2]} \tilde{q}^{[\alpha_3 \beta_3]} \tilde{q}^{[\alpha_4 \beta_4]} \tilde{q}^{[\alpha_5 \beta_5]} \tilde{q}^{[\gamma_3 \delta_3]}  
\end{align}
The Skyrmion is the large $N_c$ extension of this. It is constructed in such a way that it is an antisymmetric combination of all possible two-color labeled quarks \cite{Bolognesi_Orientifold-Skyrmion}.

To sum up, for $N_c=3$ the composites $B, \tilde{B}$ and $M_{(1,N_c-2)}$ are the standard baryons and meson, while the rest ($\tilde{B}_{N_c/2}, \tilde{B}_{even},$ $\mathcal{B}_{f_1...f_{N_c}},$ $M_{(\tilde{s},s)},$ $X_f, Y_{ff'}$) all vanish. In terms of color-structure, every one of these composites also appear in the Corrigan-Ramond extension, except $\mathcal{B}_{f_1...f_{N_c}}$, $X_f$ and $Y_{ff'}$ which are unique in this chiral setting.

One can also include the conjugate quarks $\overline{q}_{\dot{\alpha},i}, \overline{Q}^{i,f}_{\dot{\alpha}}, \overline{\tilde{q}}_{\dot{\alpha},[ij]}$ and build conjugates of all the composites we've mentioned so far, as well as some additional ones, namely $q\overline{q}$ meson-likes, arbitrarily large extensions of the tetraquark and flavor-analogues created by swapping any number of $q$'s with $Q$'s.
\begin{align*}
    &q^i\overline{q}_i \ ,  
    \qquad \tilde{q}^{[ij]} \overline{\tilde{q}}_{[ij]} \ ,
    \qquad \tilde{q}^{[i_1i_2]} \overline{\tilde{q}}_{[i_2i_3]} ... \tilde{q}^{[i_{m-1}i_{m}]} \overline{\tilde{q}}_{[i_mi_1]}\\ 
    & q^i \overline{\tilde{q}}_{[i_1i_2]} \tilde{q}^{[i_2i_3]} ... \tilde{q}^{[i_{m-1}i_{m}]} \overline{q}_{i_m} \ , 
    \qquad \tilde{q}^{[i_1i_2]} \overline{\tilde{q}}_{[i_2i_3]} ... \overline{\tilde{q}}_{[i_mi_{m+1}]} q^{i_m} q^{i_{m+1}} \ ,\\
    &+ \ \text{flavor analogues}(q^i \xrightarrow[]{} \overline{Q}^{i,f}, \overline{q}_i \xrightarrow[]{} Q_{i,f})
\end{align*}
We will consider these composites later in the text. \\

\subsection{Hadron Masses}
In this section, we will use our large $N_c$ tools to examine some of the properties and interactions of different composites in the theory. We will use diagrams and group-theoretical factors (quadratic Casimirs) to determine the masses of the composites. Afterwards, we examine interactions between composites using large $N_c$ diagram techniques.
\subsubsection{Mass: Diagrams and combinatorics} \label{section:HiddenQCD_Mass_diagrams}
\noindent First, we will use a diagrammatic approach. The goal is to examine gluon-corrections to the propagation of the hadrons with many quarks (which is every hadron except $X_f$ and $Y_{ff'}$)\\[-10pt]
\begin{align*}
    \includegraphics{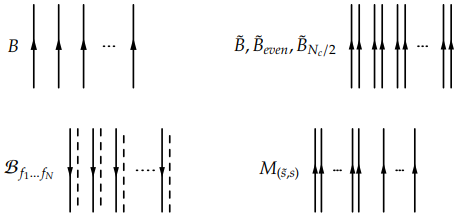}
\end{align*}

\noindent \textbf{Gluon exchanges}:
The first-order correction to $N_c$-quark diagrams come from a one-gluon exchange between two of the confined quarks. 
The possible exchanges are\\[-10pt]
\begin{align*}
    \includegraphics{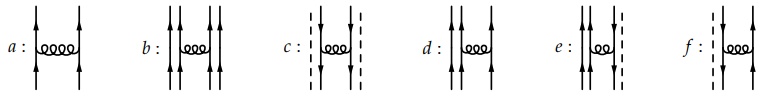}
\end{align*}
To specify what the color-flow of these diagrams look like, and then determine $N_c$-scaling, we burrow a procedure from \cite{Manohar_LargeNQCD} which is
\begin{itemize}
    \item Select a suitable vacuum-bubble
    \item Cut the quark lines twice to create the desired gluon-exchange
    \item Find the $N_c$-scaling by examining how many loops are destroyed
\end{itemize}
We now comment on each of the one-gluon exchange diagrams above. 

\noindent $a:$ The first diagram is an exchange between two fundamental quarks. Using double-line notation
\begin{align*}
    \includegraphics{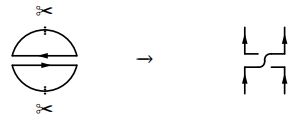}
\end{align*}
The diagram above show how to get gluon-exchange $a$ by cutting a suitable vacuum bubble and twisting the gluon-propagator. Let us count $N_c$'s: The vacuum bubble is $\sim N_c$, but cutting the quark-lines has the side-effect of destroying two color-loops $\sim 1/N_c^2$. In total this leaves us with $\sim \frac{1}{N_c}$ for the resulting diagram. Notice that our knowledge of the color-flow comes from the planar diagrams.
\\[-8pt]

\noindent $b:$ The second type of gluon exchange would be seen in the $\tilde{B}, \tilde{B}_{even}, \tilde{B}_{simple}$ baryons and the $M_{(\tilde{s},s)}$-family. A suitable diagram for analysis is a first-order correction to the $\tilde{q}^{[ij]}$ vacuum bubble
\begin{align*}
    \includegraphics{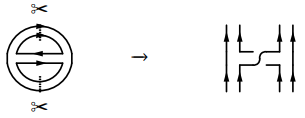}
\end{align*}
The vacuum bubble is $\sim N_c^2$ while cutting the quark-lines destroys three color-loops $\sim 1/N_c^3$. This leaves us with $\sim \frac{1}{N_c}$ in total.\\[-8pt]

\noindent $c:$ The third type of gluon-exchange only shows up in the $\mathcal{B}_{f_1...f_{N_c}}$ baryons. Again, we should cut a suitable vacuum bubble diagram
\begin{align*}
    \includegraphics{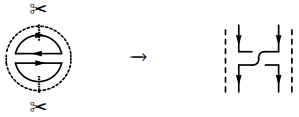}
\end{align*}
The vacuum bubble is $\sim N_c^2$ while cutting the quark-lines destroys one flavor-loop $\sim 1/N_c$ and two color-loops $\sim 1/N_c^2$. This leaves us with $\sim \frac{1}{N_c}$ in total.\\[-8pt]

\noindent $d:$ The fourth type only shows up in the $M_{(\tilde{s},s)}$ composites. Cutting
\begin{align*}
    \includegraphics{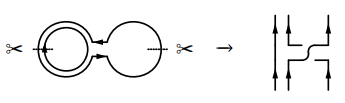}
\end{align*}
The vacuum bubble is $\sim N_c$ while cutting the quark-lines destroys both color-loops $\sim 1/N_c^2$. This leaves us with $\frac{1}{N_c}$.\\[-8pt]

\noindent $e:$ The fifth type of gluon-exchange only happens in $Y_{f,f}$\\[-10pt]
\begin{align*}
    \includegraphics{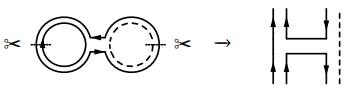}
\end{align*}
The vacuum bubble is $\sim N_c^2$ while cutting the quark-lines destroys one flavor-loop $\sim 1/N_c$ and two color-loops $\sim 1/N_c^2$. This leaves us with $\sim \frac{1}{N_c}$.\\[-8pt]

\noindent $f:$ The sixth and final type of gluon exchange only happens in $X_f$
\begin{align*}
    \includegraphics{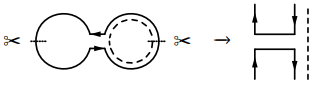}
\end{align*}
The vacuum bubble is $\sim N_c$ while cutting the quark-lines destroys one flavor-loop $\sim 1/N_c$ and one color-loop $\sim 1/N_c$. This leaves us with $\sim N_c \times \frac{1}{N_c} \times \frac{1}{N_c} = \frac{1}{N_c}$.\\

\noindent As we might have expected from an inspection of the Lagrangian, all of the one-gluon exchanges are of the same order in $N_c$. The next step in determining the mass of composites is to consider the combinatoric factors. \\

\noindent \textbf{Combinatorics and Mass}:
The combinatorics of the order $N_c$-quark composites are straightforward. They can each make $\sim N_c^2$ distinct quark pairs with each pair contributing with a $1/N_c$ gluon interaction. 
Thus, in total, the gluon interactions (and the masses of the composites) scale like $\sim \frac{1}{N_c}\times N_c^2 = N_c$.\\
The Skyrmion $\tilde{B}$ however, is a different story. 
When examining the combinatorics of the $\tilde{B}$, we have to consider a new subtlety \cite{Cherman-Cohen_The_Skyrmion_strikes_back}. 
A general one-gluon exchange changes the color-indices of a quark-pair from $ij,kl$ to $ik,jl$
\begin{align*}
    \includegraphics{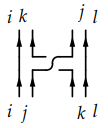}
\end{align*}
Because the Skyrmion is an antisymmetric combination of all possible two-color labeled quarks, the $ik,jl$ pair already exists in the $\tilde{B}$ composite. This violates Pauli's exclusion principle, and we must conclude that not all gluon exchanges are valid within the $\tilde{B}$ composite \cite{Cherman-Cohen_The_Skyrmion_strikes_back}.\footnote{Note that this is not a problem for $\tilde{B}_{N_c/2}, \tilde{B}_{even}$ or $\tilde{M}_{(\tilde{s},s))}$.} However, if the two quarks in the one-gluon exchange share \textit{a single} color-index, the process does not change the color of the quark pair (up to a permutation)\\
\begin{align*}
    \includegraphics{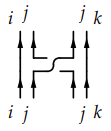}
\end{align*}
What does this mean for the combinatorics of a one-gluon exchange? When choosing the first quark, there are $\sim N_c^2$ to choose from, but for the second, there are only $\sim N_c$ quarks that share one color-index with the first. Combining these yields a total combinatoric factor of $N_c^3$.
Gluon-interaction in the Skyrmion (and the mass) is thus $\sim \frac{1}{N_c} \times N_c^3 = N_c^2$.\\
\indent For two-gluon exchanges, there are no restrictions on which quarks can participate. Therefore we have $\sim N_c^2$ choices for \textit{both} the first \textit{and} the second quark, giving a total combinatoric factor of $N_c^4$.
Because two-gluon exchanges contribute as $\frac{1}{N_c^2}$ \footnote{If we insist on our procedure of cutting diagrams, the two-gluon exchange in the diagram can be obtained from a non-planar diagram. This gives the $1/N_c^2$ factor.}, the total contribution to the Skyrmion mass is again $N_c^2$ 
\begin{align*}
    \includegraphics{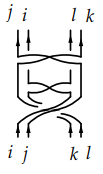}
\end{align*}
This argument can be extended to any and all classes of diagrams exclusively made of quarks in the two index antisymmetric representation \cite{Cohen-Shafer-Lebed_Roundabout}. The proof uses simplified equivalent diagrams by introducing 'gluon-roundabouts'.\\

\noindent 
The above analysis only includes two-body interactions. How does it extend to $n$-body interactions? We will use the 't Hooft baryon $B$ to argue that $n$-body interactions give the same result as two-body.
If we for instance cut a planar vacuum bubble with three gluons into three pieces, we destroy three color-loops, ending up with a factor of $N_c^{-2}$. To balance this, when you choose three quarks from the baryon, it can be done $N_c^3$ ways. Again, the total contribution is $N_c$.
\begin{align*}
    \includegraphics{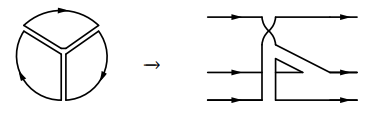}
\end{align*}
The generalization is that an $n$-body interaction scales like $N_c^{1-n}\times N_c^n=N_c$.\\

\noindent Lastly, $d$ disconnected pieces contribute with $N_c^d$. When we uncover that $M_B\sim N_c$, it is easy to identify higher order contributions with the expansion of the propagator
\begin{align}
    e^{iM_Bt} = 1 - iM_Bt - \frac{M_B^2 t^2}{2} + ...
\end{align}
These points about $n$-body interactions and multiple disconnected pieces hold for the other hadrons as well. With this in mind, we will keep just discussing two-body interactions. One might worry that this kind of diagrammatic analysis is not rigorous enough, or maybe only holds for 'heavy' quarks. To show that the results also are valid for light quarks, we will now perform an alternative set of calculations, and end up with the same results.

\noindent 
\subsubsection{Mass: Hamiltonian approach}\label{section:HiddenQCD_Mass_Casimir}
\noindent Another way of analyzing the masses of composites, is a 'semirelativistic' approach described in \cite{Buisseret_Light_baryon_masses} for light quarks.\\
Our simple baryon-Hamiltonian contains a relativistic kinetic term, and two potential terms. The one-gluon exchange potential $V_{oge}$ contains contributions from all quark-pairs, and the confining potential $V_c$ receives contributions from each quark:
\begin{align}
    H_{baryon} = T + V_{oge} + V_c =
    \sum_{i=1}^{n_q} \left[ \sqrt{\vec{p}_i^2} + a_q \sigma \left|\vec{x}_i - \vec{R}\right| \right] - \sum_{i<j=1}^{n_q} \frac{\lambda_0}{2}\frac{b_{q_1q_2}}{|\vec{x}_i - \vec{x}_j|}
    \label{eq:HiddenQCD_semirelativistic_Hamiltonian}
\end{align}
where $\sigma$ is the string tension between quarks ($\sigma$ is expected to be independent of $N_c$ at leading order \cite{Buisseret_Light_baryon_masses}), and $n_q$ is the number of quarks in the composite. The strength of both potentials depends on the \textit{representation} of the quarks that are part of the baryon:
\ytableausetup{mathmode, boxframe=normal, boxsize=2.2mm}
\begin{align}
    a_q = \frac{C_r}{C_F}, \qquad 
    b_{q_1q_2}= \frac{C_{r_1 \otimes r_2}-C_{r_1}-C_{r_2}}{N_c}
    \label{eq:HiddenQCD_a_and_b_coefficients}
\end{align}
where $C_r$ is the quadratic Casimir of the representation of a single quark in the composite, and $C_{r_1 \otimes r_2}$ is the quadratic Casimir of the representation of a pair of quarks undergoing a gluon-exchange. 
Clearly, the next step is finding the relevant quadratic Casimir operators. Using
\begin{align}
    C_r = \frac{d_{Adj}}{d_r} T_r  = \frac{(N_c+1)(N_c-1)}{d_r}T_r
\end{align}
or equivalently, if one only knows the Dynkin indices $(a_1,a_2,...,a_{N_c-1})$ of a representation,
\begin{align}
    C_r = \sum_{m=1}^{N_c-1} \left[ 
    N_c(N_c-m)m a_m + m (N_c-m) a_m^2 + \sum^{m-1}_{n=0} 2n(N_c-m)a_n a_m
    \right]
\end{align}
the relevant Casimir's are obtained and collected in Table \ref{table:QuadraticCasimirs}.\\
\ytableausetup{mathmode, boxframe=normal, boxsize=1.4mm}
\begin{table}[H]
\centering
\begin{tabular}{c|ccc}
& \cellcolor[HTML]{EFEFEF}$d_r$ & \cellcolor[HTML]{EFEFEF}$T_r$ & \cellcolor[HTML]{EFEFEF} $C_r$ \\ 
\hline \cellcolor[HTML]{EFEFEF} & & &\\[-6pt]
\cellcolor[HTML]{EFEFEF}$\ytableaushort{~}$ & $N_c$ & $\frac{1}{2}$ & $\frac{(N_c^2-1)}{2N_c}$ \\
\cellcolor[HTML]{EFEFEF} & & & \\
\cellcolor[HTML]{EFEFEF}$\mbox{Adj.}$ & $(N_c+1)(N_c-1)$ & $\frac{N_c-2}{2}$ & $N_c$ \\
\cellcolor[HTML]{EFEFEF} & & & \\
\cellcolor[HTML]{EFEFEF}$\ytableaushort{~,~}$ & $\frac{N_c(N_c-1)}{2}$ & $\frac{N_c-2}{2}$ & $\frac{(2N_c^2 - 2N_c - 4)}{2N_c}$ \\
\cellcolor[HTML]{EFEFEF} & & & \\[1pt]
\cellcolor[HTML]{EFEFEF}$\ytableaushort{~~}$ & $\frac{N_c(N_c+1)}{2}$ & $\frac{N_c+2}{2}$ & $\frac{(2N_c^2 + 2N_c - 4)}{2N_c}$ \\
\cellcolor[HTML]{EFEFEF} & & & \\
\cellcolor[HTML]{EFEFEF}$\ytableaushort{~~,~}$ & $\frac{N_c(N_c-1)(N_c+1)}{3}$ & $\frac{N_c^2-3}{2}$ & $\frac{3(N_c^2-3)}{2N_c}$ \\
\cellcolor[HTML]{EFEFEF} & & & \\[1pt]
\cellcolor[HTML]{EFEFEF}$\ytableaushort{~,~,~}$ & $\frac{N_c(N_c-1)(N_c-2)}{6}$ & $\frac{(N_c-3)(N_c-2)}{4}$ & $\frac{3(N_c^2 - 2N_c - 3)}{2N_c}$ \\
\cellcolor[HTML]{EFEFEF} & & & \\[3pt]
\cellcolor[HTML]{EFEFEF}$\ytableaushort{~~,~~}$ & $\frac{N_c^2(N_c+1)(N_c-1)}{12}$    & $\frac{(N_c+2)N_c(N_c-2)}{6}$ & $\frac{4(N_c+2)(N_c-2)}{2N_c}$ \\
\cellcolor[HTML]{EFEFEF} & & & \\[1pt]
\cellcolor[HTML]{EFEFEF}$\ytableaushort{~~,~,~}$ & $\frac{(N_c+1)N_c(N_c-1)(N_c-2)}{8}$  & $\frac{(N_c-2)(N_c^2-N_c-4)}{4}$ & $\frac{4(N_c^2 - N_c - 4)}{2N_c}$ \\
\cellcolor[HTML]{EFEFEF}{\color[HTML]{333333} } & & & \\[3pt]
\cellcolor[HTML]{EFEFEF}{\color[HTML]{333333} $\ytableaushort{~,~,~,~}$} & $\frac{N_c(N_c-1)(N_c-2)(N_c-3)}{24}$  & $\frac{(N_c-2)(N_c-3)(N_c-4)}{12}$ & $\frac{4(N_c+1)(N_c^2 -N_c-4)}{2N_c}$\\
\cellcolor[HTML]{EFEFEF} & & & \\[6pt]
\cellcolor[HTML]{EFEFEF} $r_P$ & $N_c(N_c+1)(N_c-2)$ & $\frac{(3N_c+1)(N_c-1)}{2N_c}$ & $\frac{3N_c^2 -2N_c -1}{2N_c}$\\[-2pt]
\cellcolor[HTML]{EFEFEF} & & & 
\end{tabular}
\caption{\textit{The dimension, trace-normalization and quadratic Casimirs of all relevant representations. See part {\bf b)} in the text for an explanation of why all these representations are relevant.}}
\label{table:QuadraticCasimirs}
\end{table}
$\;$

\noindent \textbf{The confining potential}: As summarized in Table \ref{table:HiddenQCD_a_coefficients}, the $a_q$-coefficients from \eqref{eq:HiddenQCD_a_and_b_coefficients} are all $\sim 1$.
\begin{table}[H]
    \centering
    \begin{tabular}{c|cl}
         & \cellcolor[HTML]{EFEFEF} $C_r/C_F \quad \;$ & \cellcolor[HTML]{EFEFEF} $\mbox{Features in}$ \\
         \hline
         \cellcolor[HTML]{EFEFEF} $\ytableaushort{~}$ & $1$ & $B,M_{(\tilde{s},s)},X_f$ \\[6pt]
         \cellcolor[HTML]{EFEFEF} $\overline{\ytableaushort{~}}$ & $1$ & $\mathcal{B}_{f_1...f_{N_c}},X_f,Y_{ff'}$ \\[6pt]
         \cellcolor[HTML]{EFEFEF} $\ytableaushort{~,~}$ & $\frac{2(N_c-2)}{(N_c-1)}$ & $\tilde{B}, \tilde{B}_{N_c/2}, \tilde{B}_{even}, M_{(\tilde{s},s)}, Y_{ff'}$
    \end{tabular}
    \caption{\textit{The color-channel-dependent coefficients of the confining potential $V_c$.}}
    \label{table:HiddenQCD_a_coefficients}
\end{table}
\noindent This means that all the composites have a contribution to their Hamiltonian proportional to the number of quarks $(\sim \sigma n_q)$, see eq.\eqref{eq:HiddenQCD_semirelativistic_Hamiltonian}.\\

\noindent \textbf{The one-gluon exchange potential}: 
Equation \eqref{eq:HiddenQCD_semirelativistic_Hamiltonian} instructs us to sum over all possible quark pairs that can undergo a one-gluon exchange. We will call the number of pairs $P$. See Table \ref{table:HiddenQCD_b_pair_counting} for a counting of pairs in each composite, and how much they each contribute to the one-gluon exchange potential.
\begin{table}[H]
    \centering
    \begin{tabular}{c|lll}
         & \cellcolor[HTML]{EFEFEF} $n_q$ & \cellcolor[HTML]{EFEFEF} $2P$ & \cellcolor[HTML]{EFEFEF} Large $N_c$ contribution to $V_{oge}$ \\
         \hline
         \cellcolor[HTML]{EFEFEF} $B$ & $N_c$ & $N_c\times(N_c-1)$ & $N_c^2b_{qq}$\\[6pt]
         \cellcolor[HTML]{EFEFEF} $\Tilde{B}_{even}$ & $N_c$ & $N_c\times(N_c-1)$ & $N_c^2b_{\tilde{q}\tilde{q}}$ \\[6pt]
         \cellcolor[HTML]{EFEFEF} $\Tilde{B}_{N_c/2}$ & $\frac{N_c}{2}$ & $\frac{N_c}{2}\times(\frac{N_c}{2}-1)$ & $N_c^2b_{\tilde{q}\tilde{q}}$ \\[6pt]
         \cellcolor[HTML]{EFEFEF} $\mathcal{B}_{f_1...f_{N_c}}$ & $N_c$ & $N_c\times(N_c-1)$ & $N_c^2b_{QQ}$\\[6pt]
         \cellcolor[HTML]{EFEFEF} $\Tilde{B}$ & $\frac{N_c(N_c-1)}{2}$ & $\frac{N_c(N_c-1)}{2} \times (\frac{N_c(N_c-1)}{2}-1) \qquad$ & $N_c^4 b_{\tilde{q}\tilde{q}}$\\[12pt]
         \cellcolor[HTML]{EFEFEF} $M_{(1,N_c-2)}$ & $1$ & $1\times(N_c-1)/2 $ & $N_cb_{\tilde{q}Q} $\\ 
         \cellcolor[HTML]{EFEFEF} & $+(N_c-1) \qquad$ & $+ (N_c-1)\times (N_c-2)$ & $+ N_c^2b_{qq}$\\[16pt]
         \cellcolor[HTML]{EFEFEF} $M_{(\tilde{s},s)}$ & $\tilde{s}$  & $\tilde{s}(\tilde{s}-1) \;$ & $\tilde{s}(\tilde{s}-1) b_{\tilde{q}\tilde{q}}$\\
         \cellcolor[HTML]{EFEFEF} & $+s$  & $+s(s-1)$ & $+ s(s-1)b_{qq}$\\
         \cellcolor[HTML]{EFEFEF} & & $+\tilde{s}s $ & $+ \tilde{s}s b_{\tilde{q}Q}$\\[16pt]
         \cellcolor[HTML]{EFEFEF} $X_f$ & $1+1$ & $1/2$ & $b_{qQ}$\\[6pt]
         \cellcolor[HTML]{EFEFEF} $Y_{f,f'}$ & $1+2$ & $3/2$ & $2b_{Q\Tilde{q}} + b_{\tilde{q}\tilde{q}}$ 
    \end{tabular}
    \caption{\textit{The different composites each have different number of quark-pairs, affecting the strength of the one-gluon potential $V_{oge}$. Notice that some composites have multiple different types of pairs. These are separated by a line-space.}}
    \label{table:HiddenQCD_b_pair_counting}
\end{table}
\noindent The next step is to insert the right $b$-coefficients into Table \ref{table:HiddenQCD_b_pair_counting}. But which ones should we use? There are six quarks-pairs, and each have multiple possible color-channels
\ytableausetup{mathmode, boxframe=normal, boxsize=2.2mm}
\begin{align}
    \ytableaushort{~} \otimes \ytableaushort{~} = \ytableaushort{~~} \oplus \ytableaushort{~,~}, &&
    &\ytableaushort{~} \otimes \overline{\ytableaushort{~}} = \mbox{Singlet} + \mbox{Adj}  \nonumber \\
    \ytableaushort{~,~} \otimes \ytableaushort{~,~} 
    = \ytableaushort{~~,~~} \oplus \ytableaushort{~~,~,~}
    \oplus \ytableaushort{~,~,~,~}, && 
    &\ytableaushort{~} \otimes \ytableaushort{~,~} = \ytableaushort{~,~,~} \oplus \ytableaushort{~~,~} \nonumber \\
    \overline{\ytableaushort{~}} \otimes \overline{\ytableaushort{~}} = \overline{\ytableaushort{~,~}} \oplus \overline{\ytableaushort{~~}}, &&
\end{align}
The sixth tensor-product deserves more space and a dimensional check:
\ytableausetup{mathmode, boxframe=normal, boxsize=3.2mm}
\begin{align}
    \ytableaushort{~,~} \otimes \overline{\ytableaushort{~}} &=
    \ytableaushort{~} \oplus
    \begin{ytableau}
    ~ & ~ \\
    ~ & ~ \\
    \none[\scalebox{0.5}{\LARGE \vdots} ] \\
    ~
    \end{ytableau}\\
    \frac{N(N-1)}{2} \times \overline{N} &= N + \frac{N(N+1)(N-2)}{2}
\end{align}
The last representation on the RHS has $N_c-1$ boxes in the first column, and two in the second. We will call this representation $r_P$ because its shape resembles that of the letter P.\\

\noindent Having enumerated the different possible color-channels, we can now calculate their individual coefficients. As an example, see the following calculation
\ytableausetup{mathmode, boxframe=normal, boxsize=1.2mm}
\begin{align}
    b_{\tilde{q}Q}=\frac{1}{N_c} C_{r_P} - C_{\bar{F}} - C_{A2} 
    &= \frac{1}{2N_c^2} \left[ 3N_c^2 -2N_c -1 -(N_c^2 - 1) - (2N_c^2 -N_c-2)\right] \nonumber \\ 
    &= \frac{1}{2N_c^2} \left[ -N_c +2\right] = +\frac{1}{N_c^2} - \frac{1}{2N_c}
\end{align}
\noindent The results are gathered in Table \ref{table:HiddenQCD_ColorChannels_b}. When considering whether a color-channel is allowed, we should check if it gives rise to attractive or repulsive interactions between the quarks in the large $N_c$ limit. This is also summarized in the table.

\ytableausetup{mathmode, boxframe=normal, boxsize=1.4mm}
\makeatletter
\def\hlinewd#1{%
\noalign{\ifnum0=`}\fi\hrule \@height #1 \futurelet
\reserved@a\@xhline}
\makeatother
\newcolumntype{?}{!{\vrule width 1pt}}
\begin{table}
\centering
\begin{tabular}{c|c|l|l}
    & \cellcolor[HTML]{EFEFEF} & \cellcolor[HTML]{EFEFEF} & \cellcolor[HTML]{EFEFEF} \\[-14pt]
    & \cellcolor[HTML]{EFEFEF}Possible Color Channels & \cellcolor[HTML]{EFEFEF}$b_{q_1q_2}$ & \cellcolor[HTML]{EFEFEF}Attractive? \\[3pt] 
    \hlinewd{1pt} \cellcolor[HTML]{EFEFEF} & & & \\[-16pt]
    \cellcolor[HTML]{EFEFEF} $\ytableaushort{~} \otimes \ytableaushort{~}$ & $\ytableaushort{~~}$  & $-\frac{1}{N_c^2} +\frac{1}{N_c}$ & $\times$ \\[2pt]
    \cline{2-4}\cellcolor[HTML]{EFEFEF} & & & \\[-16pt]
    \cellcolor[HTML]{EFEFEF} & $\ytableaushort{~,~}$  &  $-\frac{1}{N_c^2} -\frac{1}{N_c}$ & $\checkmark$ \\[6pt]
    \hlinewd{1pt} \cellcolor[HTML]{EFEFEF} & & & \\[-16pt] 
    \cellcolor[HTML]{EFEFEF} $\ytableaushort{~} \otimes \overline{\ytableaushort{~}}$ & Adj. & $+\frac{1}{N_c^2}$ & $\times$ \\[2pt] 
    \cline{2-4} \cellcolor[HTML]{EFEFEF} & & & \\[-16pt]
    \cellcolor[HTML]{EFEFEF} & Singlet & $+\frac{1}{N_c^2}-1$ & $\checkmark$  \\[2pt]
    \hlinewd{1pt} \cellcolor[HTML]{EFEFEF} & & & \\[-16pt]
    \cellcolor[HTML]{EFEFEF} $\ytableaushort{~} \otimes \ytableaushort{~,~}$ & $\ytableaushort{~~,~}$  & $-\frac{2}{N_c^2} + \frac{1}{N_c}$ & $\times$ \\[6pt]
    \cline{2-4} \cellcolor[HTML]{EFEFEF} & & & \\[-16pt]
    \cellcolor[HTML]{EFEFEF} & $\ytableaushort{~,~,~}$ & $-\frac{2}{N_c^2} -\frac{2}{N_c}$ & $\checkmark$ \\[10pt]
    \hlinewd{1pt} \cellcolor[HTML]{EFEFEF} & & & \\[-16pt]
    \cellcolor[HTML]{EFEFEF} $\ytableaushort{~,~} \otimes \ytableaushort{~,~}$  & $\ytableaushort{~~,~~}$  & $-\frac{4}{N_c^2}+\frac{2}{N_c}$ & $\times$ \\[6pt]
    \cline{2-4} \cellcolor[HTML]{EFEFEF} & & \\[-16pt]
    \cellcolor[HTML]{EFEFEF} & $\ytableaushort{~~,~,~}$ & $-\frac{4}{N_c^2}$ & $\checkmark$ \\[10pt]
    \cline{2-4} \cellcolor[HTML]{EFEFEF} & & & \\[-16pt]
    \cellcolor[HTML]{EFEFEF} & $\ytableaushort{~,~,~,~}$ & $-\frac{4}{N_c^2} - \frac{4}{N_c}$ & $\checkmark$ \\[14pt]
    \hlinewd{1pt} \cellcolor[HTML]{EFEFEF} & & & \\[-16pt]
    \cellcolor[HTML]{EFEFEF} $\ytableaushort{~,~} \otimes \overline{\ytableaushort{~}}$   & $r_P$  & $+\frac{1}{N_c^2} - \frac{1}{2N_c}$ & $\checkmark$ \\[2pt]
    \cline{2-4} \cellcolor[HTML]{EFEFEF} & & & \\[-16pt]
    \cellcolor[HTML]{EFEFEF} & $\ytableaushort{~}$  & $+ \frac{2}{N_c^2} + \frac{1}{N_c} -1$ & $\checkmark$ \\[2pt]
    \hlinewd{1pt} \cellcolor[HTML]{EFEFEF} & & & \\[-16pt]
    \cellcolor[HTML]{EFEFEF} $\overline{\ytableaushort{~}} \otimes \overline{\ytableaushort{~}}$ & $\overline{\ytableaushort{~~}}$  & $-\frac{1}{N_c^2} +\frac{1}{N_c}$ & $\times$ \\[2pt]
    \cline{2-4} \cellcolor[HTML]{EFEFEF} & & & \\[-16pt]
    \cellcolor[HTML]{EFEFEF} & $\overline{\ytableaushort{~,~}}$  & $-\frac{1}{N_c^2} -\frac{1}{N_c}$ & $\checkmark$ 
\end{tabular}
    \caption{\textit{The color-coefficients for the one-gluon-exchange potential $V_{oge}$}.}
    \label{table:HiddenQCD_ColorChannels_b}
\end{table}
\ytableausetup{mathmode, boxframe=normal, boxsize=1.2mm}
\noindent Notice that every tensor-product yields just one attractive representation except for $\ytableaushort{~,~} \otimes \ytableaushort{~,~}$ and $\ytableaushort{~,~} \otimes \overline{\ytableaushort{~}}$. 
Furthermore, the resulting color-channels (such as $\ \ytableaushort{~~,~,~}$ and $\ytableaushort{~,~,~,~} \ $) have different large $N_c$ behaviour, so the distinction is important.\\ 
\indent First, we will look at the choice between $r_P$ and $\ytableaushort{~}$. 
In this case, the $b$-coefficient for $\ytableaushort{~}$ is $\sim 1$, and simply dominates over the coefficient for $r_P$ which is $\sim 1/N_c$.\\
\indent Next, the choice between $\ytableaushort{~~,~,~}$ and $\ytableaushort{~,~,~,~}$. This is in fact what we treated in the combinatorics of the diagrammatic analysis. For $\tilde{B}_{N_c/2},\Tilde{B}_{even}$ and $M_{(\tilde{s},s)}$, the $\ytableaushort{~,~,~,~}$ representation works just fine, and it also naturally dominates over $\ytableaushort{~~,~,~}$ (their $b$-coefficients are $\sim 1/N_c$ and $\sim 1/N_c^2$ respectively). However, the Skyrmion $\tilde{B}$ is constructed in such a way that a one-gluon exchange can only happen between quarks that share one color index. This is exactly what the $\ytableaushort{~~,~,~}$-representation does - two of the colors are symmetric as apparent from the Young tableu.\\

\noindent Finally, we can insert the attractive $b_{q_1q_2}$ coefficients, and calculate $V_{oge}$ for the different hadrons. The results are gathered in Table \ref{table:HiddenQCD_b_inserted}.
\begin{table}[]
    \centering
    \begin{tabular}{c|ll}
         & \cellcolor[HTML]{EFEFEF} $n_q$ & \cellcolor[HTML]{EFEFEF} Large $N_c$ contribution to $V_{oge}$ \\
         \hline
         \cellcolor[HTML]{EFEFEF} $B$ & $N_c$ & $N_c$\\[+6pt]
         \cellcolor[HTML]{EFEFEF} $\tilde{B}_{N_c/2}$ & $N_c/2$ &  $N_c$ \\[6pt]
         \cellcolor[HTML]{EFEFEF} $\Tilde{B}_{even}$ & $\frac{N_c}{2}$ &  $N_c$ \\[6pt]
         \cellcolor[HTML]{EFEFEF} $\mathcal{B}_{f_1...f_{N_c}}$ & $N_c$ &  $N_c$\\[6pt]
         \cellcolor[HTML]{EFEFEF} $\Tilde{B}$ & $\frac{N_c(N_c-1)}{2}$ & $N_c^2$\\[16pt]
         \cellcolor[HTML]{EFEFEF} $M_1$ & $1+(N_c-2) \qquad$  & $N_c$\\[16pt] 
         \cellcolor[HTML]{EFEFEF} $M_{(\tilde{s},s)}$ & $\tilde{s} + s$  & $\frac{\tilde{s}s + \tilde{s}(\tilde{s}-1) + s(s-1)}{N} \sim N_c$\\[16pt]
         \cellcolor[HTML]{EFEFEF} $X_f$ & $1+1$ & $1$\\[6pt]
         \cellcolor[HTML]{EFEFEF} $Y_{f,f'}$ & $1+2$ &  $1$ 
    \end{tabular}
    \caption{\textit{Final contributions to the one-gluon-potential $V_{oge}$ after calculating number of pairs, and then weighing each pair with their respective quadratic Casimir.}}
    \label{table:HiddenQCD_b_inserted}
\end{table}
\noindent To conclude on the one-gluon-potential: Once again, each hadron gets a contribution proportional to their number of quarks.\\

\noindent \textbf{Bounds on hadron mass based on} $a_q$ \textbf{and} $b_{q_1q_2}$:
We have now seen how both the potentials of the Hamiltonian are of order $n_q$. Next, with a little help from \cite{Buisseret_Light_baryon_masses}, we can set an upper and lower bound for the masses of our composites
\begin{align}
    n_q \inf_{\ket{\psi}} \langle{\psi} | \sqrt{\vec{p}^2} + \frac{n_q-1}{2} \left[ \frac{a_q}{n_q}r + \frac{b_{q_1q_2}}{r} \right] | {\psi} \rangle
    \leq M_{composite} \leq
    \sqrt{4a_q (Q n_q- b_{q_1 q_2} P^{3/2})}
    \label{eq:HiddenQCD_Mass_UpperLowerBound}
\end{align}
where $Q \sim n_q$ is the band number of the considered state in a harmonic oscillator picture. Inserting into eq. \eqref{eq:HiddenQCD_Mass_UpperLowerBound} and taking the large $N_c$ limit, we again arrive at the masses with the same $N_c$ scaling as $\sim n_q$.\\
\indent Before moving on, we should note a detail about the masses of $\tilde{B}_{N_c/2}$ and $\tilde{B}_{even}$. Because they are symmetric in their gauge-functions under exchange of two quarks, they receive a contribution from Fermi zero temperature pressure (which maximally scales like $\sim N_c^{4/3}$) \cite{Bolognesi_Orientifold-Skyrmion}.\\

\subsection{Hadron Interactions}
\subsubsection{Interactions: Scattering diagrams}
\noindent Following a diagrammatic analysis similar to Witten \cite{Witten_Baryons}, we can get a heuristic idea about how scattering between the different hadrons ($B,\tilde{B}_{N_c/2},\Tilde{B}_{even}, \Tilde{B},\mathcal{B}_{f_1...f_{N_c}},M_{(\tilde{s},s)},X_f,Y_{ff'}$) scale with $N_c$\footnote{It should be noted that this analysis is at fixed velocity and not fixed momentum because many of the hadrons have $N_c$-scaling masses.}.\\
\indent The idea is to examine all possible exchanges between the hadrons, using the gluon exchanges from section \ref{section:HiddenQCD_Mass_diagrams}. 
To evaluate the final effect of these diagrams, we have to develop a procedure to determine the associated combinatoric factors. 
To establish this procedure, we should study some examples:\\

\noindent \underline{Exchanges between $B$ and $X_f$:} How might these interact? They could exchange a quark, gluon or both. Let's draw all possible diagrams up to one-gluon exchange\\[-4pt]
\begin{align*}
    \includegraphics{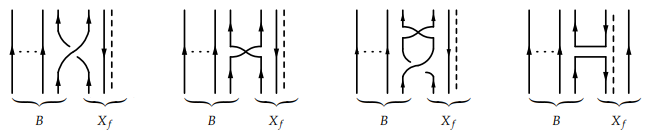}
\end{align*}
When evaluating these diagrams, we need to take into account non-trivial combinatoric factors. To highlight this, we look closer at the quark-quark exchange channel with one gluon (the second and third diagrams).\\
First, assign a color $i$ to the in-going quark from composite with lowest $n_q$, in this case $X_f$. This gives a combinatoric factor of $1$ (there is \textit{one} fundamental quark in $X_f$ to choose from). Follow the color-flow, and carry the color to where it ends
\begin{align*}
    \includegraphics{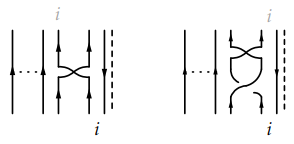}
\end{align*}
Next, the color of all in-going quarks should match their outgoing color for the hadron to be gauge invariant. We should assign the $i$ to the outgoing quark from $X_f$, and follow the color-flow again (this time backwards)
\begin{align*}
    \includegraphics{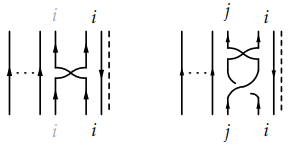}
\end{align*}
At this point the diagrams differ. The colors of the first diagram are fully determined giving a combinatoric factor of $1$ (there is only \textit{one} quark in $B$ with the color $i$), while there is an undetermined color index $j$ in the second diagram giving a combinatoric factor $N_c$ (there are $N_c$ quarks in $B$ with an arbitrary color). 
Thus, the diagrams (with combinatoric factors) contribute $1/N_c$ and $1$ respectively, so $\mathcal{M}(B,X_f)_{scattering} \sim 1$. This is strong enough to affect the $X_f$'s, but not the 't Hooft baryons.\\
\indent If we had instead examined the $q$-$Q$ channel (the fourth diagram above), we would have found that diagram to contribute $\sim 1/N_c$. It is a trend that \textit{cross-quarktype channels are suppressed} - although sometimes they are still the most dominant channels.\\

\noindent \underline{$\tilde{q}$ combinatoric factors:} Because the Skyrmions $\Tilde{B}$ have $N_c(N_c-1)$ quarks, they can give a larger combinatoric factor than eg. $B_{N_c/2}$ with $N_c/2$ quarks. Lets see an example of this by examining their respective interaction with $\mathcal{B}_{f_1...f_{N_c}}$:
\begin{align*}
    \includegraphics{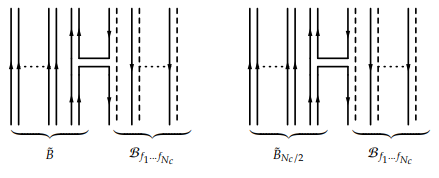}
\end{align*}
In both diagrams, the interacting quark has one determined color $i$, and one free color. For the interaction with $\tilde{B}$, this means a combinatoric factor of $N_c$ (there are $N_c$ quarks in $\tilde{B}$ with color $i$). For $\tilde{B}_{N_c/2}$, the combinatoric factor is $1$ (there is just \textit{one} quark in $\tilde{B}_{N_c/2}$ with color $i$).
We conclude that $\mathcal{M}(\tilde{B},\mathcal{B}_{f_1...f_{N_c}})_{scattering} \sim N_c$ (strong enough to affect $\mathcal{B}_{f_1...f_{N_c}}$ but not $\tilde{B}$), while $\mathcal{M}(\tilde{B}_{N_c/2},\mathcal{B}_{f_1...f_{N_c}})_{scattering} \sim 1$ (not strong enough to affect either of the interacting hadrons).\\
\indent 
Next, compare the scattering amplitudes for pairs $(B,X_f)$ and $(\tilde{B}_{N_c/2},Y_{ff'})$. Naively, one might think that the amplitudes should be the same - the $B,\tilde{B}_{N_c/2}$'s are somewhat similar in structure, the $X,Y$'s are similar, and both pairs can exchange a quark of the same type\\[-10pt]
\begin{align*}
    \includegraphics{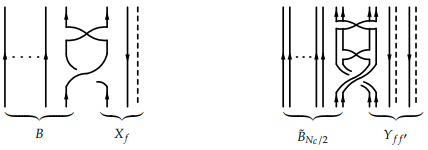}
\end{align*}
We've already seen that $\mathcal{M}(B,X_f)_{scattering} \sim 1$ for the first diagram. In the second diagram, the two gluons give $1/N_c^2$, there is $1$ quark to choose from in $Y$, and $N_c/2$ quarks to choose from in $\tilde{B}_{N_c/2}$. In total, we get $\mathcal{M}(\tilde{B}_{N_c/2},Y_{ff'})_{scattering} \sim 1/N_c$. The $\tilde{B}_{N_c/2}$'s do not affect the $Y_{ff'}$'s even though they have quark content in common.\\
The two amplitudes differ because of the two index antisymmetric quarks $\tilde{q}^{[ij]}$. A quark $\tilde{q}^{[ij]}$ exchange requires \textit{two} gluon exchanges for the colors to be independent of each other. This allows for large combinatoric factors when $\tilde{B}$ is involved, but also serves to suppress quark exchanges in $\tilde{B}_{N_c/2}, \tilde{B}_{even},M_{(\tilde{s},s)}$.\\ 

\noindent \underline{$M_{(\tilde{s},s)}$ counting:} Because they have a variable number of quarks $q^i$ and $\tilde{q}^{[ij]}$, the $M_{(\tilde{s},s)}$ exchanges require a little more of a delicate counting. Let's examine the scattering amplitudes for the pair $(M_{(\tilde{s},s)},B)$
\begin{align*}
    \includegraphics{Scattering_Diagrams_5.png}
\end{align*}
The first diagram, with combinatoric factors, contributes $\sim \frac{1}{N_c}\times \tilde{s} \times 2 = \frac{\tilde{s}}{N_c}$ (there are $\tilde{s}$ quarks to choose from, and one quark in $B$ with matching colors).
The second diagram, with combinatoric factors, contributes $\sim \frac{1}{N_c} \times s \times N = s$. The second diagram dominates, and we get $\mathcal{M}(M_{(\tilde{s},s)},B)_{scattering} \sim s$.\\

\noindent \underline{General procedure:} For two composites $v,w$ with $n_q(v)\leq n_q(w)$, the general procedure is
\begin{itemize}
    \item Assign colors to the ingoing quark $q_1$ from $w$. Follow the color-flow and assign colors to wherever they end up.
    \item Assign the same colors to the outgoing quark $q_1$ from $w$, and follow the color-flow backwards to again ensure that the colors are consistent.
    \item Determine the two combinatoric factors by asking, a: How many quarks could I have chosen for step 1? b: How many quarks in $v$ have colors that are consistent with those determined in step 1 and 2.
\end{itemize}
\noindent Now we have the tools to examine interactions between any hadron.
Exhausting every possible one-gluon interaction (or two-gluon when two index antisymmetric quarks $\tilde{q}^{[ij]}$ are exchanged), we arrive at table \ref{table:HiddenQCD_ScatteringAmpltudes_1}:
\begin{table}[H]
    \centering
    \begin{tabular}{l|ccccccccc}
         & \cellcolor[HTML]{EFEFEF} $\Tilde{B}$ & \cellcolor[HTML]{EFEFEF} $\tilde{B}_{N_c/2}$ & \cellcolor[HTML]{EFEFEF} $\tilde{B}_{even}$ & \cellcolor[HTML]{EFEFEF} $M_{(\tilde{s},s)}$ & \cellcolor[HTML]{EFEFEF} $B$ & \cellcolor[HTML]{EFEFEF} $\mathcal{B}_{f_1...f_{N_c}}$ & \cellcolor[HTML]{EFEFEF} $X_f$ & \cellcolor[HTML]{EFEFEF} $Y_{f,f'}$  \\
         \hline
         \cellcolor[HTML]{EFEFEF} & & & & & & & & \\[-10pt]
         \cellcolor[HTML]{EFEFEF} $\tilde{B}$  
         & $N_c^2$ & $N_c$ & $N_c$ & $N_c$ & $N_c$ & $N_c$ & $1$ & $1$ \\
         \cellcolor[HTML]{EFEFEF} & & & & & & & & \\[-8pt]
         \cellcolor[HTML]{EFEFEF} $\tilde{B}_{N_c/2}$  
         & $N_c$ & $1$ & $1$ & $1$ & $1$ & $1$ & $1/N_c$ & $1/N_c$ \\
         \cellcolor[HTML]{EFEFEF} & & & & & & & & \\[-8pt]
         \cellcolor[HTML]{EFEFEF} $\tilde{B}_{even}$  
         & $N_c$ & $1$ & $1$ & $1$ & $1$ & $1$ & $1/N_c$ & $1/N_c$ \\
         \cellcolor[HTML]{EFEFEF} & & & & & & & & \\[-8pt]
         \cellcolor[HTML]{EFEFEF} $M_{(\tilde{s},s)}$  
         & $N_c$ & $1$ & $1$ & $N_c$ & $r$ & $1$ & $1/N_c$ & $1/N_c$ \\
         \cellcolor[HTML]{EFEFEF} & & & & & & & & \\[-8pt]        
         \cellcolor[HTML]{EFEFEF} $B$  
         & $N_c$ & $1$ & $1$ & $r$ & $N_c$ & $1$ & $1$ & $1/N_c$ \\
                  \cellcolor[HTML]{EFEFEF} & & & & & & & & \\[-8pt]
         \cellcolor[HTML]{EFEFEF} $\mathcal{B}_{f_1...f_{N_c}}$
         & $N_c$ & $1$ & $1$ & $1$ & $1$ & $N_c$ & $1$ & $1$ \\
                  \cellcolor[HTML]{EFEFEF} & & & & & & & & \\[-8pt]
         \cellcolor[HTML]{EFEFEF} $X_f$  
         & $1$ & $1/N_c$ & $1/N_c$ & $1/N_c$ & $1$ & $1$ & $1$ & $1$ \\
         \cellcolor[HTML]{EFEFEF} & & & & & & & & \\[-8pt]
         \cellcolor[HTML]{EFEFEF} $Y_{f,f'}$  
         & $1$ & $1/N_c$ & $1/N_c$ & $1/N_c$ & $1/N_c$ & $1$ & $1$ & $1$ \\[+4pt]
         \end{tabular}
    \caption{\textit{Scattering amplitudes between the different hadrons. Notice that if the two hadrons have no quark content in common, the interaction can be suppressed.}}
    \label{table:HiddenQCD_ScatteringAmpltudes_1}
\end{table}
\noindent The scattering amplitudes should be compared to the masses of the involved hadrons. This is done in table \ref{table:HiddenQCD_ScatteringAmpltudes_2}:
\begin{table}[H]
    \centering
    \includegraphics[scale=0.5]{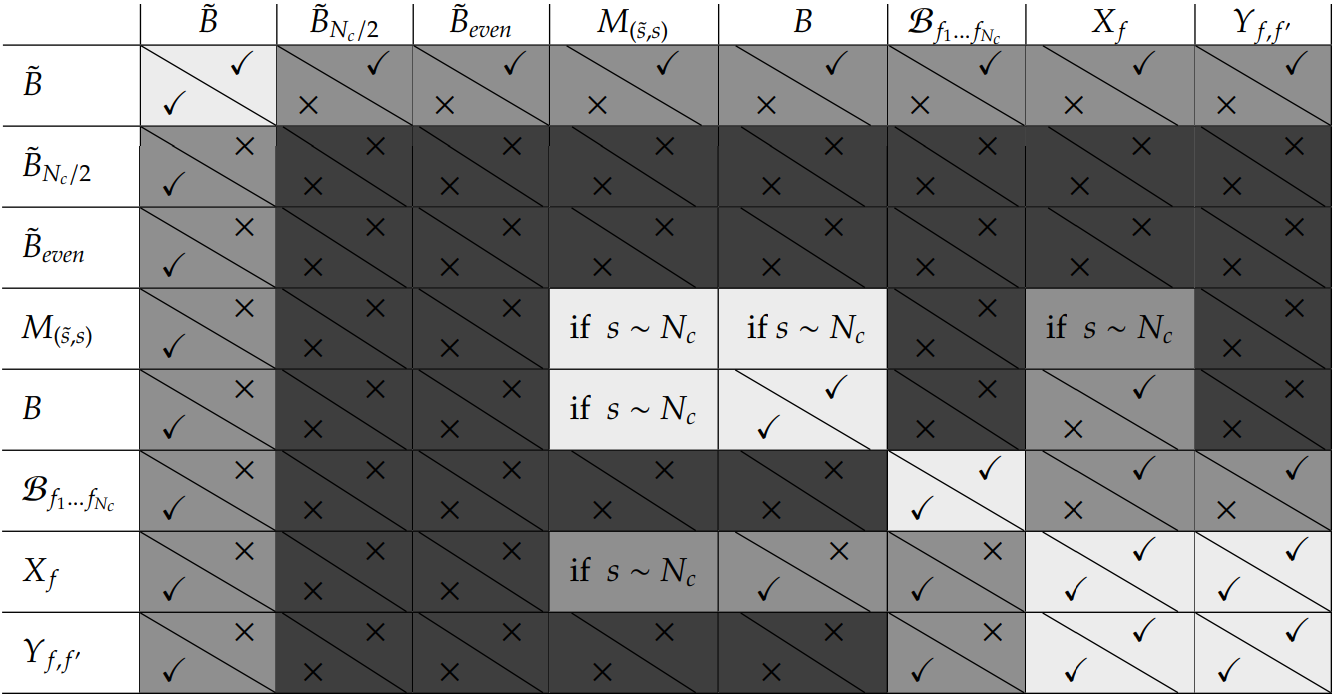}
    \caption{\textit{Qualitative analysis of scattering amplitudes. If the amplitude is strong enough to affect the motion of one of the involved hadrons, we write a checkmark in the direction of that hadron.}}
    \label{table:HiddenQCD_ScatteringAmpltudes_2}
\end{table}
\noindent We can make a few interesting observations from table \ref{table:HiddenQCD_ScatteringAmpltudes_2}. 
\begin{itemize}
    \item The $\tilde{B}$ only scatters with itself, but affects the motion of everything else around. The $\tilde{B}_{N_c/2}$ and $\tilde{B}_{even}$ on the other hand affect nothing, and are only scattered by $\tilde{B}$.
    \item The usual  meson $M_{(1,N_c-2)}$ and the 't Hooft baryon $B$ scatter with each other. This might be seen as an attractive feature of the large $N_c$ limit because it resembles what happens in the $N_c=3$ theory.
    \item The $X_f$ and $Y_{ff'}$'s scatter with each other, and their motion is affected by $\mathcal{B}_{f_1...f_{N_c}}$. The $B$'s can also push around the $X_f$'s because they both feature fundamental quarks. 
\end{itemize}
The general pattern is that scattering amplitudes between hadrons follow a mass- and quark-type hierarchy: 
Hadrons with $M \sim N_c^2$ interact with all other hadrons, and are only scattered only by themselves. Hadrons with $M \sim N_c$ scatter with each other if they have a sufficient amount of quark content in common. Hadrons with $M \sim 1$ scatter among themselves, and are scattered by higher mass hadrons if they have sufficient quark content in common. 
For fundamental and antifundamental quarks, is sufficient number of quarks is $\sim N_c$, and for two index quarks it is $\sim N_c^2$.\\

\subsubsection{Interactions: Correlation functions}
The $X_f$ and $Y_{ff'}$ composites have $n_q \sim 1$, so we can use correlation function analysis to examine them.
Before calculating any correlation functions, note that insertions of $X_f$ and $Y_{ff'}$ on quark lines are quite limited. They must appear pairwise. The possible insertions are
\begin{align*}
    \includegraphics{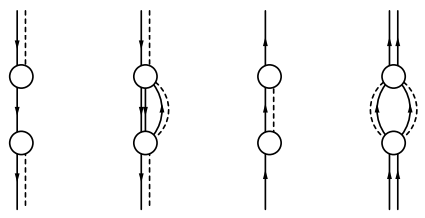}
\end{align*}
Notice that the two insertions of $Y_{ff'}$ on an antifundamental line yield an extra color-loop, but also breaks up a potential flavor-loop (remember the arrows on the flavor-lines follow those of the quark color-lines).\\

\noindent $\mathbf{Y_{f,f'}}$: It is only possible to draw \textit{even} $n_Y$-point functions of $Y_{f,f'}$.
The leading order contributions to the generating functional from $Y_{f,f'}$ are of order $N_c^{1+n_Y/2}$, and they look something like this:
\begin{align*}
    \includegraphics{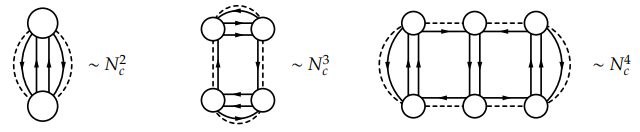}
\end{align*}
Note that instead of forming loops, the flavors are simply propagating between hadrons (remember that the flavor-lines also carry arrows with the same direction as their associated color line).
Adding sources, we learn that we should include a factor of $N_c$ in the definition of $\hat{Y}_{f,f'}$ if we want the two-point function to be of order $1$. Thus, a general $n_Y$ point function goes like:
\begin{align}
    &\langle Y(x)Y(y) \rangle = \frac{1}{N_c^2} \frac{\delta}{\delta J(x)} \frac{\delta}{\delta J(y)} \log(Z[J]_{J\equiv 0}) \sim 1\\
    &\langle \underbrace{Y\cdots Y}_{n_Y} \rangle \sim \frac{1}{N_c^{n_Y}} \times N_c^{1+n_Y/2} = N_c^{1-n_Y/2}
\end{align}
As a result of this behaviour, the $Y$'s by themselves are stable, free particles ($\langle YYYY \rangle \sim N_c^{-1}$). 
Furthermore, because the operator is single-trace, and the theory is asymptotically free, we expect the $N_c\xrightarrow[]{}\infty$ world to feature an infinite tower of single-particles states 
\cite{Witten_Baryons}.\\

\noindent $\mathbf{X_f}$: The fact that the $X_f$'s consists of two different quark types, means that they also have to appear in pairs. The leading order contributions to $\log(Z[J])$ comes from processes like:
\begin{align*}
    \includegraphics{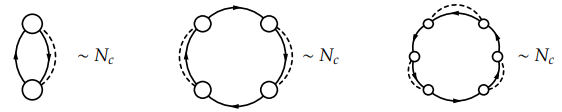}
\end{align*}
Like the standard meson operator, the diagrams contributing to $n_X$-point functions are always $\sim N_c$. Consequently, they have to be normalized by a factor $\sqrt{N_c}$, and a general $n_X$-point function is:
\begin{align}
    \langle \underbrace{X\cdots X}_{n_x} \rangle \sim \frac{1}{N_c^{n_X/2}} \times N_c = N_c^{1-n_X/2}
\end{align}
This is exactly similar to the $Y$ composite, and the conclusions that one can draw from it are the same.\\

\noindent \textbf{Mixing X's and Y's}: It is easy to create diagrams with both $X$, and $Y$ operators. The general structure can be constructed by starting with the skeleton of a $\langle YY \rangle$ diagram. Then insert $n_X$ $X$'s two at a time on an arbitrary $Q$ line, and add $n_Y$ $Y's$ two at a time, directly connecting to the previous $YY$ pair. Because the $X$'s do not change the $N_c$-counting, the resulting diagram contributes $N_c^{1+n_Y/2}$
\begin{align*}
    \includegraphics{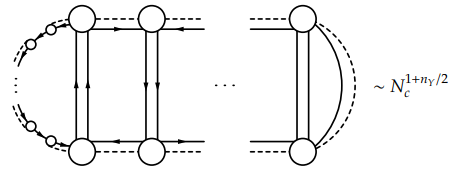}
\end{align*}
Now, an arbitrary $n$-point function is:
\begin{align}
    \langle \underbrace{X\cdots X}_{n_x} \ \underbrace{Y\cdots Y}_{n_Y} \rangle \sim \frac{1}{N_c^{n_X/2+n_Y }} \times N_c^{1+n_Y/2} = N_c^{1-n_X/2-n_Y/2} 
\end{align}
The mixed correlation function of highest order in $N_c$ is $\langle XXYY \rangle \sim N_c^{-1}$, so interactions are suppressed.\\

\noindent Do the $X$ and $Y$ mix with the glueballs? The glueballs can be inserted arbitrarily on available gluon-lines without changing the value of a diagram. And because of planarity, there are available gluon-lines everywhere. All in all then, each glueball adds a factor of $N_c^{-n_G}$ in the correlation functions, because of their normalization
\begin{align}
    \langle \underbrace{X\cdots X}_{n_x} \
    \underbrace{Y\cdots Y}_{n_Y} \
    \underbrace{G\cdots G}_{n_G}\rangle 
    \sim \frac{1}{N_c^{n_X/2 +n_Y +n_G }} \times N_c^{1+n_Y/2} = N_c^{1-n_X/2-n_Y/2-n_G} 
\end{align}
We conclude that interactions with glueballs are also suppressed.

\subsubsection{Interactions: Diagrams for additional processes}
The gauge structure of our composites also allow for nontrivial processes although many of them are suppressed. They all conserve the two $U(1)$ symmetries from Table \ref{tab:content} as they should.\\ 

\noindent \textbf{$X_f$ and $Y_{ff'}$ insertions}: The $X_f$ and $Y_{ff'}$'s can be inserted on the baryon lines. They have to be inserted two at a time, and they each bring normalization factors $\sqrt{N_c}$, and $N_c$ respectively for $X_f$ and $Y_{ff'}$. The color-flow of the baryon lines thus allows for the creation of an $X_f\overline{X}^f$ pair (or $Y_{ff'}\overline{Y}^{ff'}$ pair), or scattering between a baryon and $X_f$ (or $Y_{ff'}$). Some examples are
\begin{align*}
    \includegraphics{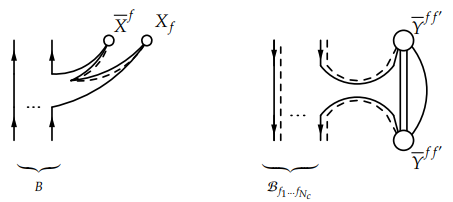}
\end{align*}
The amplitudes for a baryon ($B,\mathcal{B}_{f_1...f_{N_c}}$) to scatter with $X_f$ or create an $X_f\overline{X}^f$ pair, are the same
\begin{align}
    \mathcal{M} \sim \underbrace{N_c^{-1}}_{\text{Normalization factors}} \times \underbrace{N_c}_{\text{Combinatorics}} = 1
\end{align}
Creating a pair of particles would of course also require the baryon to transition to a lower energy-state.
The amplitudes for a baryon ($\mathcal{B}_{f_1...f_{N_c}},\tilde{B},\tilde{B}_{even},\tilde{B}_{N_c/2}$) to scatter with $Y_{ff'}$ or create a $Y_{ff'}\overline{Y}^{ff'}$ pair are also the same, but they depend on which baryon one chooses.\\ \\
For $\tilde{B}_{even},\tilde{B}_{N_c/2}$, the amplitude is
\begin{align}
    \mathcal{M} \sim \underbrace{N_c^{-2}}_{\text{Normalization factors}} \times \underbrace{N_c}_{\text{Combinatorics}} = \frac{1}{N_c}
\end{align}
For $\mathcal{B}_{f_1...f_{N_c}}$, the amplitude is
\begin{align}
    \mathcal{M} \sim \underbrace{N_c^{-2}}_{\text{Normalization factors}} \times \underbrace{N_c}_{\text{Color loop}} \times \underbrace{N_c}_{\text{Combinatorics}} = 1
\end{align}
For $\tilde{B}$, the amplitude is
\begin{align}
    \mathcal{M} \sim \underbrace{N_c^{-2}}_{\text{Normalization factors}}  \times \underbrace{N_c^2}_{\text{Combinatorics}} = 1
\end{align}
These results are compatible with Table \ref{table:HiddenQCD_ScatteringAmpltudes_2}.\\

\noindent \textbf{Baryonium States}:
Next, as an analogue of baryonium states in the 't Hooft limit \cite{Witten_Baryons,Liu_Baryonium}, a single $\mathcal{B}_{f_1...f_{N_c}}$ may combine with $B$ or $\tilde{B}_{N_c/2}$ to form a 'baryonium' bound-state $B\mathcal{B}_{f_1...f_{N_c}}$ or $\tilde{B}_{N_c/2}\mathcal{B}_{f_1...f_{N_c}}$. First consider the $B \mathcal{B}_{f_1\ldots f_{N_c}}$ state. With the inclusion of combinatoric factors, interactions between the fundamental and antifundamental quarks inside this state are $\sim N_c$
\begin{align*}
    \includegraphics{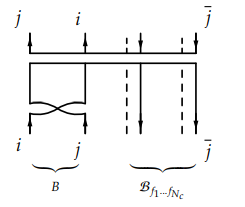}
\end{align*}
\ytableausetup{mathmode, boxframe=normal, boxsize=2.8mm}
It is possible for $B\mathcal{B}_{f_1...f_{N_c}}$ to decay via successive emissions of $X_f$'s. See Feynman diagram below. After one such emission, we will call the resulting state $B\mathcal{B}_{f_1...f_{N_c-1}}(1)$.
In this state, the remaining $(N_c-1)$ quarks $\epsilon_{i_1\ldots i_{N_c}} q^{i_1}\ldots q^{i_{N_c-1}}$ form an antifundamental representation, and the remaining $(N_c-1)$ flavored quarks $\epsilon^{i_1 \ldots i_{N_c}} Q_{i_1,f_1} \ldots Q_{i_{N_c-1},f_{N_c-1}}$ form a fundamental representation. In this sense, such a state behaves like a $q^i \Bar{q}_i$ operator - it is capable of producing color-loops, and should be normalized accordingly.

From this state, another meson may be emitted, and then another, and so on. A general baryonium state obtained by $C$ emissions is then $B\mathcal{B}_{f_1...f_{N_c-C}}(C)$. We can write such a decay as 
\begin{align}
    B\mathcal{B}_{f_1...f_{N_c-C}}(C) \xrightarrow{} B\mathcal{B}_{f_1...f_{N_c-(C+1)}}(C+1) + X_f
\end{align}
The following diagram represents the first process in which $B \mathcal{B}_{f_1\ldots f_{N_c}}$ emits the first meson $X_f$
\begin{align*}
    \includegraphics{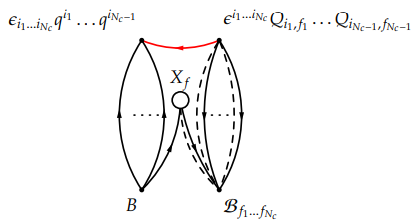}
\end{align*}
Here, the colored line is not a quark-line. Rather, it represents the fact the the antifundamental $\epsilon_{i_1\ldots i_{N_c}} q^{i_1}\ldots q^{i_{N_c-1}}$ and fundamental $\epsilon^{i_1 \ldots i_{N_c}} Q_{i_1,f_1} \ldots Q_{i_{N_c-1},f_{N_c-1}}$ states combine to form a gauge-singlet.
The amplitude for an arbitrary $X_f$ emission is
\begin{align}
    \mathcal{M}_{B\mathcal{B}(C) \xrightarrow[]{} B\mathcal{B}(C+1)} 
    \sim \underbrace{\frac{1}{N_c^{C/2}} \times \frac{1}{N_c^{(C+1)/2}} \times \frac{1}{N_c^{1/2}}}_{B\tilde{B}, X_f \text{ Normalization factors}} \times \underbrace{(N_c-C)}_{\text{Combinatoric factor}} \times \underbrace{N_c^C}_{\text{Color loops}}
    = \frac{N_c-C}{N_c}
\end{align}
The amplitude starts at $\sim 1$ when the bound state consists of $2N_c$ quarks, and ends at $\sim \frac{1}{N_c}$ when there are two quarks left.\\

In the other baryonium state $\tilde{B}_{N_c/2}\mathcal{B}_{f_1...f_{N_c}}$ there again exists interactions between the $Q$ and $\tilde{q}$ which are $\sim N_c$. This baryonium state decays via successive emissions of $Y_{ff'}$'s. The following diagram represents the first emission
\begin{align*}
    \includegraphics{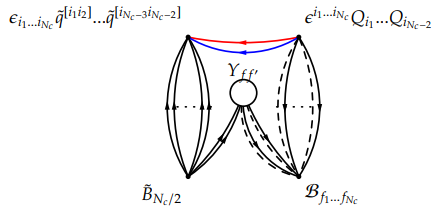}
\end{align*}
The amplitude for this process is:
\begin{align}
    \mathcal{M}_{\tilde{B}_{N_c/2}\mathcal{B}(C) \xrightarrow[]{}  \tilde{B}_{N_c/2}\mathcal{B}(C+1)} 
    & \sim \underbrace{\frac{1}{N_c^{C}} \times \frac{1}{N_c^{C+1}} \times \frac{1}{N_c}}_{B\tilde{B},Y_{ff'} \text{ Normalization factors}} \times \underbrace{\frac{N_c}{2}-C}_{\text{Combinatoric factor}} \times \underbrace{N_c^{2C}}_{\text{Color loops}} \nonumber \\
   & = \frac{N_c/2-C}{N_c^2}
\end{align} 
The amplitude starts at $\sim \frac{1}{N_c}$ when there are $\frac{3}{2}N_c$ bound quarks and ends at $\sim \frac{1}{N_c^2}$ when there are three quarks left. The $\tilde{B}_{N_c/2-C}\mathcal{B}_{f_1...f_{N_c-C}}(C)$ states are thus narrower than the $B\mathcal{B}_{f_1...f_{N_c-C}}(C)$ states by a factor of $N_c$.\\

\noindent 
\textbf{A Non-Trivial Scattering Process}:
We now consider the following non-trivial scattering process 
\begin{align}
     M_{(\tilde{s},s)} + Y_{ff'} \xrightarrow[]{} 
     2X_f + M_{(\tilde{s}+1,s-2)}
\end{align}
Diagrammatically it is
\begin{align*}
   \centering
    \includegraphics[scale=0.4]{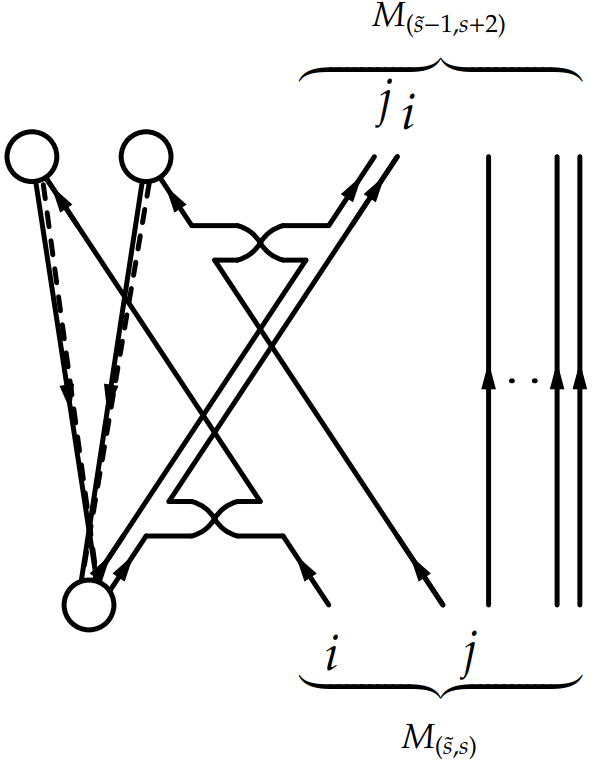}
\end{align*}
The amplitude is
\begin{align}
    \mathcal{M}_{(\tilde{s},s)\xrightarrow[]{}(\tilde{s}+1,s-2)} \sim \frac{1}{N_c^2} \times \frac{1}{N_c^2} \times N_c^2 \times \tilde{s} = \frac{\tilde{s}}{N_c^2}
\end{align}
Maximally this process scales as $\sim \frac{1}{N_c}$, which is suppressed compared to the scaling of the mass of $M_{(\tilde{s},s)}$.

\subsection{Flavor Structure in the Hadronic Spectrum}
So far, we have studied the composite spectrum of the theory without including the conjugated quarks $\overline{q}_{{\dot\alpha},i}, \overline{\tilde{q}}_{{\dot\alpha},[ij]}, \overline{Q}^{i,f}_{\dot\alpha}$.
We will now, for the sake of completeness, consider broadly the spectrum, correlation functions and interactions when they are included.\\

\noindent \textbf{Flavor in the chiral large $N_c$ extension}: The antifundamental quarks $Q_{i,f}$ have an $SU(N_c-3)$ flavor-index, while the fundamental quarks $q^i$, and two-index antisymmetric quarks $\tilde{q}^{[ij]}$ do not. 
How does this flavor-index contribute in interactions between hadrons?
First, any intermediate state will be filled by gluons, two-index antisymmetric quarks $\tilde{q}$ and quarks with a flavor index $Q_f$, because they all have free energy scaling as $\sim N_c^2$. 
Second, the flavor index can form loops. Flavor loops can only be created by flavor-contracted operators such as $T_{f}^{\phantom{f}f} = Q_{i,f} \tilde{q}^{[ij]} \overline{\tilde{q}}_{[jk]} \overline{Q}^{k,f}$. Third, operators can have free flavor indices, which do not alter the scaling of any given diagram.

When considering the full spectrum, there are groups of hadrons that have all gauge indices contracted in an identical manner, but for which their flavor indices differ (consider eg. $X_f=q^i Q_{i,f}$ versus $X = q^i \overline{q}_i$). In this section we will divide the spectrum into composite operators that contain no $Q_{f}$ quarks and all the remaining ones that contain at least one $Q_{f}$ quark. The former set therefore contains no flavor indices (free or contracted) while the latter set contains at least one flavor index. In the case where the latter set contains multiple $Q_{f}$ fields the flavor indices can be either free or contracted. The former set of composite operators that contains no $Q_{f}$ quarks will be called flavorless operators while the latter set that contains at least one $Q_{f}$ quark will be called flavored operators. Note that this distinction implies that the two flavor singlet operators $T = \Bar{q}_{i} \tilde{q}^{[ij]} \overline{\tilde{q}}_{[jk]} q^{k}$ and $T_{f}^{\phantom{f}f} = Q_{i,f} \tilde{q}^{[ij]} \overline{\tilde{q}}_{[jk]} \overline{Q}^{k,f}$ do not belong to the same group. According to our definition the former composite operator is flavorless while the latter is flavored. We will now study the interactions of the flavorless and flavored hadrons. 

In the full spectrum, gauge singlet flavorless operators composed of $n_q$ fields have multiple gauge singlet flavored analogues also composed of $n_q$ fields. The spin may in general differ between the various operators. Consider as an example the following four-quark operator 
\begin{align}
    T = q^i \overline{\tilde{q}}_{[ij]} \tilde{q}^{[jk]} \overline{q}_k
\end{align}
There exists three flavored four-quark analogues of this operator, one for each possible $q \xrightarrow{} Q$ swap.
\begin{align*}
    T_f = q^i \overline{\tilde{q}}_{[ij]} \tilde{q}^{[jk]} Q_{k,f}\\
    T^f_{f'} = \overline{Q}^{i,f} \overline{\tilde{q}}_{[ij]} \tilde{q}^{[jk]} Q_{k,f'}
\end{align*}
These three operators comprise the full set of four-quark gauge singlet flavored operators (without any gluon operators).
Notice that the spin of the flavorless $T$ and flavored operators $T_f, T^f_{f'}$ are typically not the same. Also note that the operator $T^{f}_{f'}$ decomposes into a flavor singlet $T^f_f$ and flavor adjoint of the $SU(N_c-3)$ flavor group.

We can use three-point functions to highlight how diagrams with flavor-loops differ from diagrams without. Compare $\langle TTT \rangle$-diagrams with $\langle T^f_{f}T^{f'}_{f'}T^{f''}_{f''} \rangle$-diagrams 
\begin{align*}
    \includegraphics{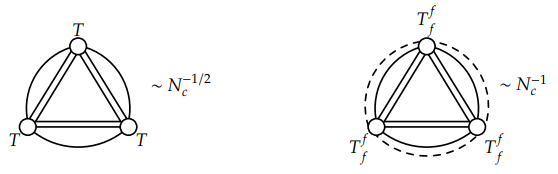}
\end{align*}
First the flavor-loop provides a factor of $\sim N_c$ to a diagram. Second any flavor-contracted operator like $T^f_f$ will have a higher normalization factor by a factor of $\sqrt{N_c}$ compared to the flavorless analogue $T$. In combination this yields an additional  $N_c \times N_c^{-3/2} = N_c^{-1/2}$ suppression of the second diagram compared to the first.

Generally, any $n$-point function of $T^f_f$ is $\sim N_c^{(2-n)/2}$ suppressed compared to an $n$-point function of $T$.\\

Consider now composite operators in which all gauge indices are contracted in a non-trivial way. In other words the composite operator is not composed of a product of gauge invariant operators. In addition we also take the operator to not have any gauge indices contracted by $\epsilon_{i_1...i_{N_c}}$ of the gauge group $SU(N_c)$. This implies that the one-index quarks $q^i,Q_{i,f}$ and their conjugated $\overline{q}_i,\overline{Q}^{i,f}$ can appear at most twice in the operator. 
In addition to these restrictions, we will temporarily ignore operators composed entirely of two-index antisymmetric quarks $\tilde{q}$.\\
\textbf{Flavorless hadrons}: What is left of the hadron spectrum is quite limited. Before discussing their flavored analogues, we will first describe the flavorless hadrons with two one-index quarks. There are infinitely many of them, but they are easy to enumerate
\begin{equation}
\label{eq:FlavorlessHadrons_two_q}
\begin{aligned}
    X &= q^i \overline{q}_i \\
    Y &= q^i q^j \overline{\tilde{q}}_{[ij]} \\
    T &= q^i \overline{\tilde{q}}_{[ij]} \tilde{q}^{[jk]} \overline{q}_k\\
    P &= q^i \overline{\tilde{q}}_{[ij]} \tilde{q}^{[jk]} \overline{\tilde{q}}_{[kl]} q^k\\
    H &= q^i \overline{\tilde{q}}_{[ij]} \tilde{q}^{[jk]} \overline{\tilde{q}}_{[kl]} \tilde{q}^{[lh]} \overline{q}_h\\
    ...
\end{aligned}
\end{equation}
Allowed $n$-point functions involving any combination of the above hadrons all scale exactly the same in $N_c$.
The reason is as follows: In the leading order diagrams, there is one fundamental quark loop running on the boundary $\sim N_c$. Each operator insertion then creates $\tilde{s}$ additional color loops (one for each $\tilde{q}$) $\sim N_c^{\tilde{s}}$, implying a $N_c^{(1+2\tilde{s})/2}$ normalization factor. Thus, each operator gives a factor of $N_c^{\tilde{s}}\times N_c^{-(1+2\tilde{s})/2} = N_c^{-1/2}$.
In total then, $n$-point functions with flavorless operators in Eq. \ref{eq:FlavorlessHadrons_two_q} scale like $\sim  N_c^1 (N_c^{-1/2})^n = N_c^{1-n/2}$.

Some $n$-point functions vanish. Two examples are $\langle YYY \rangle$ and $\langle PYX \rangle$. This is so since in general, a $\tilde{q}$-line started by one operator must be ended by another. Because of the color structure of the various quarks, this also enforces the restriction that fermionic operators cannot form odd $n$-point functions.\\

Motivated by the fact that all $n$-point functions scale exactly the same, we will now construct some simplified diagrams that will quickly visualize how the $n$-point functions of the above described hadrons are organized.
Any three-point function is represented by a triangle. The lines of the triangle represent the quark loop and the three vertices represent the three operators. An example is\\[-20pt]
\begin{align*}
    \includegraphics{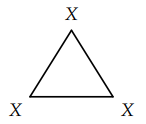}
\end{align*}
To keep track of the two-index antisymmetric quarks, each vertex additionally produces $\tilde{s}$ lines internally in the triangle 
\begin{align*}
    \includegraphics{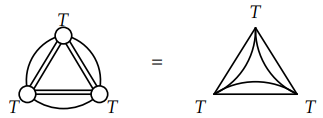}
\end{align*}
With this setup, all the three point functions with $X,Y,T,P,H$ are\\[-8pt]
\begin{align*}
   \centering
    \includegraphics[scale=0.5]{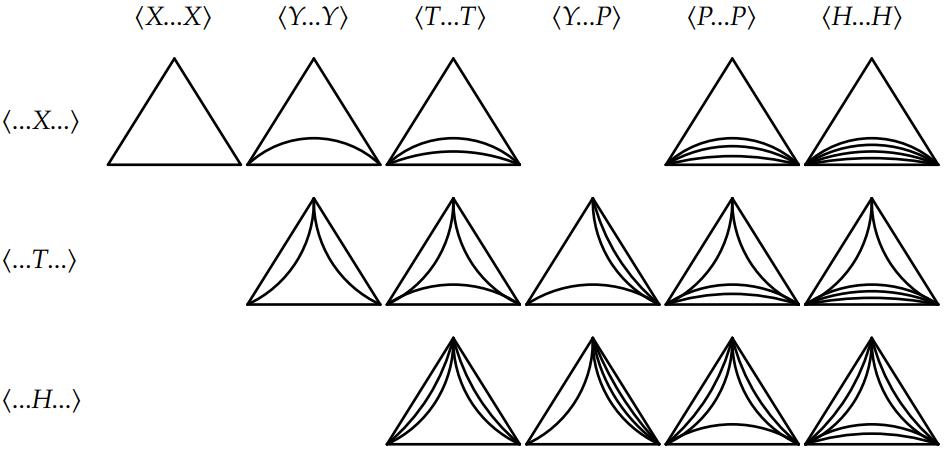}
\end{align*}
These all scale like $\sim N_c^{-1/2}$.
Four-point functions are squares scaling like $\sim N_c^{-1}$, five-points are pentagons scaling like $\sim N_c^{-3/2}$ etc. At this point we have a solid handle on this group of flavorless operators. All their $n$-point functions scale the same, and we can quickly find out what $n$-point functions are allowed by gauge structure.\\

\textbf{Flavored hadrons}: For each of the above flavorless diagrams, there are a number of equivalent diagrams in the flavored sector - mixing flavorless operators with flavored ones, and flavored ones with other flavored ones. 
In our simplified diagrams, these can be created from eg. the flavorless triangles by replacing any of the solid outer lines with dashed flavor-lines (the dashed line is a shorthand for the $Q_{i,f}$ propagator). An example is the diagram for $\langle X_f X^f X \rangle$
\begin{align*}
    \includegraphics{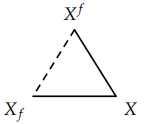}
\end{align*}
This allows us to quickly enumerate, for any three-point function of operators appearing in Eq. \ref{eq:FlavorlessHadrons_two_q}, \textit{all} flavored analogues. Keep in mind that the inside of the diagram is irrelevant when comparing flavored diagrams to flavorless ones\\[-8pt]
\begin{align*}
   \centering
    \includegraphics[scale=0.5]{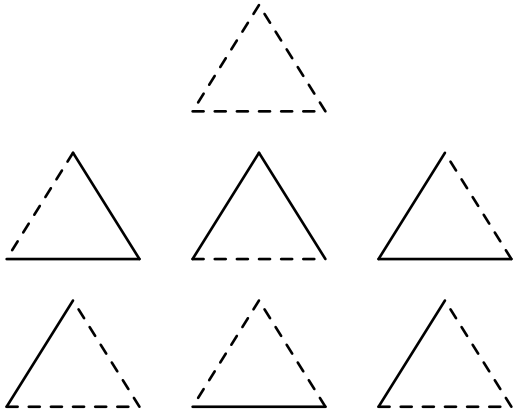}
\end{align*}
These diagrams tell us about how the flavorless sector interacts with the flavored sector. 
First, we see that the flavor-operators must appear in pairs or larger groups. Second, the scaling of a generic diagram is not altered by substitution of any number of the solid lines with dashed lines. However, if all $n$ operators are flavor-contracted (like $T^f_f$), they form a flavor loop and the diagram is suppressed by a factor of $N_c^{(2-n)/2}$ as we noted earlier. Third, for flavorless three-point functions with \textit{one} operator such as $\langle TTT \rangle$, there are three flavored analogues.
For \textit{two} different operators, fx. $\langle XYY \rangle$, there are five flavored analogues. For flavorless three-point functions with \textit{three} different operators, fx. $\langle YTP \rangle$, all of the above seven triangles are unique flavored analogues of the flavorless triangle.
This can be generalized to arbitrary $n$-point functions. In general, depending on how many different operators are involved, an $n$-point function will have between $n_{-} -1$ and $n_+-1$ flavored analogues where
\begin{align}
    n_- = \begin{cases}
        \frac{2n^2 + 16}{8}, \quad n \ \text{even}\\
        \frac{2n^2 + 14}{8}, \quad n \ \text{odd}
    \end{cases},
    \qquad \qquad n_+ = 2^n
\end{align}
\\[-6pt]

\noindent \textbf{Hadrons made exclusively of two-index quarks}: Now we come back to the operators made exclusively of two-index quarks. They are also easy to enumerate
\begin{align*}
    L_1 &= \tilde{q}^{[ij]} \overline{\tilde{q}}_{[ji]}\\
    L_2 &= \tilde{q}^{[ij]} \overline{\tilde{q}}_{[jk]} \tilde{q}^{[kl]} \overline{\tilde{q}}_{[li]}\\
    L_2 &= \tilde{q}^{[ij]} \overline{\tilde{q}}_{[jk]} \tilde{q}^{[kl]} \overline{\tilde{q}}_{[lh]} \tilde{q}^{[hr]} \overline{\tilde{q}}_{[ri]}\\
    ...
\end{align*}
In \cite{Cohen-Lebed_Tetraquarks_QCD_AS} the authors studied $n$-point functions of $L_2$ operators so we will move forward and study $n$-point functions when including the $X,Y,T,P,H...$ operators. When combined with the $X,Y,T,P,H...$ operators the $L_i$, $i=1,\ldots$ operators should be inserted inside the quark loop where the two-index quarks are.
Each such insertion costs a factor of $N_c^{-1}$. This is because any $L_i$ insertion creates $i-1$ extra color loops, but also comes with a normalization factor of $N_c^i$.
In terms of our simplified diagrams, we may add the $L_i$ operators as points that consume / produce $2i$ internal lines. An example is
\begin{align*}
    \includegraphics{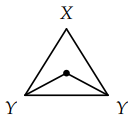}
\end{align*}
Again we can diagrammatically enumerate every possible $n$-point function.
Using $X,Y,T$, all four-point functions with just one $L_i$ are given by the diagrams\\[-8pt]
\begin{align*}
   \centering
    \includegraphics[scale=0.5]{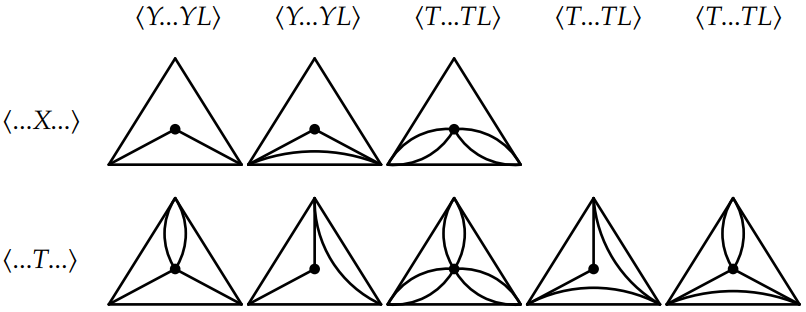}
\end{align*}
These all scale like $\sim N_c^{-3/2}$.
What happens inside the triangles does not affect our flavor considerations - it is still true that any of the above diagrams have flavor-analogues that scale exactly the same.\\

Only the flavor-loops in the above analysis utilize the fact that $f$ is a flavor-index with $\sim N_c$ flavors, and they end up being suppressed. Nevertheless, we have sketched a diagrammatic program for charting a subsection of the full hadronic spectrum, and shown that interactions between the flavorless and flavored sectors follow simple patterns.
One-flavor index operators begin interacting with their flavorless analogues in three-point functions $\sim N_c^{-1/2}$. 
Two-flavor index operators begin interacting with their flavor-analogues in four-point functions $\sim N_c^{-1}$.
A notable exception to the suppression of mixing between the flavorless and flavored sectors are interactions such as $\langle X^f_f L_1 \rangle \sim 1$.

\section{Discussion}\label{conclusions}

We set out to study QCD in a large $N_c$ extension which is intermediate between the renowned 't Hooft and Orientifold extensions. Using standard large $N_c$ techniques, we have systematically examined properties and interactions of multiple composites in the hadronic spectrum in the chiral extension of QCD.

First, we examined the purely left-handed and right-handed part of the hadronic spectrum. Here, 
$B, \tilde{B}$ and $M_{(1,N_c-2)}$ are the standard baryons and meson, while the rest ($\tilde{B}_{N_c/2}, \tilde{B}_{even},$ $\mathcal{B}_{f_1...f_{N_c}},$ $M_{(\tilde{s},s)},$ $X_f, Y_{ff'}$) all vanish at $N_c = 3$. 

Using a diagrammatic analysis and group theoretical factors, we determined the $N_c$ scaling of the different hadron masses to be proportional to the number of constituent quarks. 
Next, a study of simple gluon exchanges between hadrons showed that scattering amplitudes between hadrons follow a mass- and quark-type hierarchy: 
Hadrons with masses $M \sim N_c^2$ interact with all other hadrons, and are scattered only by themselves. Hadrons with masses $M \sim N_c$ interact with each other if they have a sufficient amount of quark content in common. Hadrons with masses $M \sim 1$ interact among themselves, and are scattered by higher mass hadrons if they have sufficient quark content in common. 
Only even $n$-point functions of $X_f$ and $Y_{ff'}$ are non-trivial, and they are $N_c^{1+n/2}$ suppressed. However, $X_f$ and $Y_{ff'}$ play crucial roles in other processes. They appear in the decay of two distinct baryonium states. The amplitude for such a decay depends on how many quarks are in the bound states, and ranges from $\sim N_c^0$ to $\sim N_c^{-1}$ for the first baryonium, and from $N_c^{-1}$ to $N_c^{-2}$ for the second baryonium. We also considered a non-trivial scattering process between $M_{(\tilde{s},s)}$ and $Y_{ff'}$, and found it to be at least $N_c^{-1}$ suppressed.

Finally we considered composites made of a mix of the left- and right-handed quarks. 
To break down the daunting task of analyzing the complete spectrum, we divided it into manageable groups. The hadron operators that do not make use of $\epsilon_{i_1...i_{N_c}}$, and are composed of two fundamental flavorless quarks, and any number of two-index antisymmetric quarks, are particularly easy to understand. Correlation functions involving $n_1$ of these operators all scale like $\sim N_c^{(2-n_1)/2}$. 
Next, hadron operators that are composed of $i$ pairs of two-index quarks are also easy to understand. Their $n_2$-point functions scale like $\sim N_c^{-n_2}$. 
Interactions between the two above mentioned groups of hadrons scale like $N_c^{(2-n_1-2n_2)/2}$. 
Next, by performing swaps of the type $q \xrightarrow[]{} Q$, flavored analogues of the above mentioned operators can be created. When analyzing $n$-point functions involving a mix of flavorless and flavored operators, the flavorless ones must appear in pairs. This means that mixing of the flavored and flavorless sectors begin in three-point diagrams which are $N_c^{-1/2}$ suppressed.
Generally, mixing diagrams obtained by this procedure to form $n$-point functions have the same scaling as the flavorless ones, so at $n\geq 3$, there are many diagrams mixing the flavorless and flavored sectors. In particular, for any $n$, there are $2^n-2$ different diagrams mixing flavorless and flavored operators.\\

For possible future research directions we note that it would be interesting to study this chiral extension in a holographic setting. Holographic QCD has been under intense investigations in the last 20 years \cite{Evans:Holography1, Evans:Holography2, Evans:Holography3, Kiritsis:Holography_p1, Kiritsis:Holography_p2,Hoyos-Karch_CR-baryons-and-holography}. Holography can be used to study, among numerous other topics, chiral symmetry breaking \cite{Gudnason:Holography_ChiralSymmetryBreaking}, nuclear physics with a solitonic approach \cite{Gudnason:Holography_Soliton,Gudnason:Holography_NuclearPhysics}, and several additional topics using the Veneziano limit of QCD to construct holographic models (V-QCD)  \cite{Jarvinen:Holography_VQCD1,Jarvinen:Holography_VQCD2,Jarvinen:Holography_VQCD3,Alho-Jarvinen-Kiritsis:FiniteTemperature_VQCD}. Quite recently QCD in the 't Hooft large $N_c$ limit has also been under investigations using bootstrap techniques \cite{Bootstrap1,Bootstrap2,Bootstrap3}. It would be highly interesting to do similar bootstrap studies of the chiral large $N_c$ extension investigated here. In other words we believe our investigations open up for many new and exciting research projects.

\newpage
\appendix
\section{Group Theory}\label{app:group}
We here discuss the properties of the representations entering the large $N_c$ extension. These are the fundamental representation $(F)$ and its complex conjugate representation $(\bar{F})$, the adjoint representation $(A)$, the two index antisymmetric representation $(2A)$ and its complex conjugate representation $(\overline{2A})$. 

{\bf First consider the fundamental representation $F$:} An $SU(N)$ transformation and an element $U_F$ of $SU(N)$ is an $N\times N$ matrix written as 
\begin{align}
q \rightarrow q' = U_F q \ , \qquad U_F = 1+ i \alpha_a T^a_F + O(\alpha^2) \ , \qquad a=1,\ldots,N^2-1 \ , \qquad \alpha_a \in \mathbb{R}
\end{align}
where the generators $T_F^a$ are traceless $\text{Tr} \ T^a _F= 0$ and Hermitian $\left( T^a_F \right)^{\dagger} = T^a_F$ since the group we are considering is $SU(N)$. Hermiticity of the generators implies $(T_F^{a})^* = ( T_F^{a})^T $. A general $N \times N$ Hermitian matrix has $N$ independent real diagonal entries and $2 \frac{N(N-1)}{2}$ independent real entries in the upper right (or lower left) triangle. The number of linearly independent Hermitian matrices that are traceless is therefore
\begin{align}
N + 2 \frac{N(N-1)}{2} -1 = N^2-1
\end{align}
Hence $a=1,\ldots,N^2-1$. The commutator of two generators in the fundamental representation is antihermitian and traceless. It can therefore be written as a linear combination of the generators in the fundamental representation with imaginary coefficients 
\begin{align}\label{eq:}
\left[T^a_F,T^b_F \right] = i f^{ab}_{\phantom{ab}c} T^c_F 
\end{align}
These are the Lie algebra commutation relations. The structure constants of the Lie algebra $f^{ab}_{\phantom{ab}c}  = - f^{ba}_{\phantom{ba}c} $ are all real and antisymmetric in the first two indices. From the identity 
\begin{align}
\left[ T^a_F, \left[ T^b_F,T^c_F \right] \right] + \left[ T^b_F, \left[ T^c_F,T^a_F \right] \right] + \left[ T^c_F , \left[ T^a_F,T^b_F \right] \right]  = 0 
\end{align}
we find by inserting the Lie algebra commutation relations
\begin{align}
 \left[ f^{ad}_{\phantom{ad} e} f^{bc}_{\phantom{bc} d} + f^{bd}_{\phantom{bd} e} f^{ca}_{\phantom{ca} d} + f^{cd}_{\phantom{cd} e} f^{ab}_{\phantom{ab} d}  \right] T^e_F = 0
\end{align}
Since the generators are all linearly independent this is only possible provided
\begin{align}\label{eq:Jacobi}
f^{ad}_{\phantom{ad} e} f^{bc}_{\phantom{bc} d} + f^{bd}_{\phantom{bd} e} f^{ca}_{\phantom{ca} d} + f^{cd}_{\phantom{cd} e} f^{ab}_{\phantom{ab} d} = 0
\end{align}
This is the Jacobi identity for the structure constants. The actual values of the structure constants are basis dependent. We arbitrarily change the basis of the Lie algebra by acting with a general invertible transformation on the generators
\begin{align}\label{eq:changeT}
T^a_F &\rightarrow T'^a_F = P^{a}_{\phantom{a} b} T^b_F 
\end{align}
by some invertible matrix $P$. Then in order to preserve the form of the Lie algebra this induces a transformation on the structure constants
\begin{align}
f^{ab}_{\phantom{ab}c} & \rightarrow f'^{ab}_{\phantom{'ab}c} = P^{a}_{\phantom{a}a'} P^{b}_{\phantom{b}b'} f^{a'b'}_{\phantom{a'b'}c'} \left( P^{-1} \right)^{c'}_{\phantom{c'}c}
\end{align}
So the values of the structure constants are basis dependent. We now introduce the trace form of the fundamental generators\footnote{The associated form for the generators in the adjoint representation is known as the Killing form.}
\begin{align}
K_F^{ab} &= \text{Tr} \ T_F^a T_F^b 
\end{align}
The trace form is a symmetric matrix due to the cyclic property of the trace. It is also a real matrix since the generators are Hermitian $(T_F^a)^* = (T_F^a)^T$. Finally it is also positive definite which can be seen as follows. Take any real non-zero vector $x_a \neq 0$ so that $x_a T^a_F \neq 0$ is also a non-zero Hermitian matrix. Consider then
\begin{align}
x_a K_F^{ab} x_b & = \text{Tr} \ \left( x_a T^a_F \right) \left( x_b T^b_F \right) = \text{Tr} \ \left( x_a T^a_F \right)^{\dagger} \left( x_b T^b_F \right)  = \sum_{i,j}^{N} \left| \left( x_a T^a_F \right)^{j}_{\phantom{j}i}  \right|^2 > 0 
\end{align}
So the trace form is a positive definite matrix which implies that its inverse matrix exists. This we denote by $(K_F)_{ab}$ with
\begin{align}
(K_F)_{ac} K_F^{cb} = \delta_{a}^{\phantom{c}b}
\end{align}
Since the trace form is positive definite it defines an inner product which in turn induces a metric. So we can think of $K_F^{ab}$ as a metric that we can use to raise adjoint indices and we can think of $(K_F)_{ab}$ as an inverse metric that we can use to lower adjoint indices. So one could for instance go ahead and define new structure constants 
\begin{align}
\tilde{f}^{abc} = f^{ab}_{\phantom{ab}d} K_F^{dc}
\end{align}
and using the trace form and the Lie algebra commutation relations we can write them as
\begin{align}
\tilde{f}^{abc} &=  f^{ab}_{\phantom{ab}d} K_F^{dc} = - i \text{Tr} \ \left[ T^a_F, T^b_F \right] T^c_F = - i \text{Tr} \ T^a_F \left[ T^b_F, T^c_F \right]
\end{align}
We see that that these new structure constants are completely antisymmetric in all three indices. However it is possible to simplify the situation even further. Remember that the trace form is positive definite so it is possible to diagonalize it. Consider how the fundamental generators transformed under a general change of basis in Eq. \ref{eq:changeT}. This induces a transformation of the trace form as 
\begin{align} 
K_F^{ab} &\rightarrow K'^{ab}_F = P^{a}_{\phantom{a}a'} K_F^{a'b'} P^{T \ b}_{b'} 
\end{align}
It is possible now to commit to a particular convenient basis of the Lie algebra in which the trace form simplifies. Since the trace form is real and symmetric it can be diagonalized by an orthogonal matrix $O$. Now pick the change of basis matrix $P$ to be this particular orthogonal matrix $P=O$ so that the trace form in this basis is diagonal
\begin{align}
OO^T &= 1  \\
T_F^a & \rightarrow O^{a}_{\phantom{a} a'} T^{a'}_{F}  \\\
K_F^{ab} & \rightarrow K'^{ab}_{F} = O^{a}_{\phantom{a}a'} K_F^{a'b'} O_{b'}^{T\ b}  = D^{ab}_F  
\end{align}
where $D_F^{ab}$ is a diagonal matrix. Since the trace form is positive definite its diagonal elements (eigenvalues) are positive. Recalling that we still have the freedom to rescale each generator in the fundamental representation the trace form in this diagonal basis is just proportional to the identity
\begin{align}
D_F^{ab} & = \text{Tr} \ T^a_F  T^b_F =  T_F \delta^{ab} \ , \qquad (D_F)_{ab}  = \frac{1}{T_F} \delta_{ab}
\end{align}
The positive number $T_F$ will be referred to as the trace normalization factor for the fundamental representation. As we will see once this is nailed to a specific value the normalization of the generators in the other representations will be automatically decided as well. 

Importantly in this basis in which the trace form is proportional to the identity we will use the identity $\delta^{ab}$ to raise adjoint identities and $\delta_{ab}$ to lower adjoint identities. In other words adjoint indices can just all be written as upper or lower indices and we don't have to worry about them. We will work in this basis from now on and so an element of $SU(N)$ is written as
\begin{align}
U_F &= 1+ i \alpha^a T^a_F + O(\alpha^2) \ , \qquad a=1,\ldots,N^2-1 \ , \qquad \alpha_a \in \mathbb{R}
\end{align}
while the Lie algebra is
\begin{align}\label{eq:}
\left[T^a_F,T^b_F \right] = i f^{abc} T^c_F \ , \qquad f^{abc} = - \frac{i}{T_F} \text{Tr} \ \left[ T_F^a , T_F^b \right] T_F^c
\end{align}
with the structure constants being antisymmetric in all its three indices. The generators in the fundamental representation satisfy the following important completeness identity 
\begin{align}\label{eq:completeness}
\left( T_F^a \right)^{i}_{\phantom{i}j} \left( T_F^a \right)^{k}_{\phantom{k}l} &=  T_F \left( \delta^{i}_{\phantom{i}l} \delta^{k}_{\phantom{k}j} - \frac{1}{N} \delta^{i}_{\phantom{i}j} \delta^{k}_{\phantom{k}l}  \right) 
\end{align}
This identity will be useful to us and we will now give a proof of it. Any $N \times N$ complex matrix $M_0$ can be written as
\begin{align}
M = M_0 \mathbb{I} + M^a T^a_F 
\end{align}
for some complex coefficients $M_0$ and $M^a$. We can project out the first coefficient $M_0$ by taking the trace on both sides of the equation and we can project out the coefficients $M^a$ by first multiplying with $T^b_F$ and then taking the trace. We then find
\begin{align}
M_0 = \frac{1}{N} \text{Tr} \ M \ , \qquad M^a = \frac{1}{T_F} \text{Tr} \ MT^a_F
\end{align}
and can write 
\begin{align}
M^{i}_{\phantom{i}j} = \frac{1}{N} M^{k}_{\phantom{k}k} \delta^{i}_{\phantom{i}j} + \frac{1}{T_F} M^{l}_{\phantom{l}k} \left( T_F^a \right)^{k}_{\phantom{k}l} \left( T^a_F \right)^{i}_{\phantom{i}j}
\end{align}
Shuffling things around we can write this as
\begin{align}
\left[ \frac{1}{T_F} \left( T^a_F \right)^{i}_{\phantom{i}j} \left( T^a_F \right)^{k}_{\phantom{k}l} -\delta^{i}_{\phantom{i}l} \delta^{k}_{\phantom{k}j} + \frac{1}{N} \delta^{i}_{\phantom{i}j} \delta^{k}_{\phantom{k}l}  \right] M^{l}_{\phantom{l}k} = 0 
\end{align}
Since this must be true for any complex $N \times N$ matrix $M$ the square bracket must be zero and hence the above identity for the generators $T^a_F$ follows. With the completeness relation in hand it follows immediately that the combination $T_F^a T_F^a$ is proportional to the identity
\begin{align}
T_F^a T_F^a = C_F \mathbb{I} \ , \qquad C_F = (2T_F) \frac{N^2-1}{2N}
\end{align}
We refer to $C_F$ as the quadratic Casimir of the fundamental representation. Lastly we note that the fundamental representation is complex for all $N\geq 3$. Its complex conjugate representation $\bar{F}$ is given by 
\begin{align}
Q \rightarrow U_{\bar{F}} Q \ , \qquad T^a_{\bar{F}} & = - \left(T^a_F  \right)^* \ , \qquad  \left[ T^a_{\bar{F}}, T^b_{\bar{F}} \right] = i f^{abc}  T^c_{\bar{F}}\ , \qquad U_{\bar{F}} = 1 + i \alpha^a T^a_{\bar{F}} + O(\alpha^2)
\end{align}

{\bf Second consider the adjoint representation $A$:} We define the generators in the adjoint representation via the structure constants as
\begin{align}
\left( T_A^b \right)^{ac} &=  i f^{abc}
\end{align}
The generators are $(N^2-1 )\times (N^2 -1)$ matrices and so the dimension of the adjoint representation is $N^2-1$. From the Jacobi identity Eq. \ref{eq:Jacobi} we can explicitly see that the generators in the adjoint representation satisfy
\begin{align}
\left[T^a_A , T^b_A \right] =i  f^{abc}T^c_A  
\end{align}
and therefore furnish a representation of the Lie algebra. An $SU(N)$ transformation in the adjoint representation is 
\begin{align}
A_{\mu} \rightarrow  A_{\mu}' = U_A A_{\mu}  \ , \qquad U_A = 1 + i \alpha^a T^a_A + O(\alpha^2) 
\end{align}
or viewed as a tensor
\begin{align}
A_{\mu} \rightarrow A_{\mu}' = U_F A_{\mu} U_F^{\dagger} \ , \qquad U_F = 1 + i \alpha^a T_F^a + O(\alpha^2)
\end{align}
The gauge field is expanded in the generators in the fundamental representation
\begin{align}
\left( A_{\mu} \right)^{i}_{\phantom{i}j} &= A_{\mu}^{a} \left( T_F^a \right)^{i}_{\phantom{i}j} 
\end{align}
Using the completeness relation Eq. \ref{eq:completeness} we find that the Killing form and the associated trace normalization factor for the adjoint representation explicitly are
\begin{align}
\text{Tr} \ T_A^a T_A^b = T_A \delta^{ab} \ , \qquad T_A = (2T_F)N 
\end{align}
As promised the Killing form is proportional to the identity and is fixed once $T_F$ is fixed. From this we find the quadratic Casimir for the adjoint representation
\begin{align}
T_A^a T_A^a = C_A \mathbb{I} \ , \qquad C_A = T_A = (2T_F)N
\end{align}
For the adjoint representation the trace normalization factor and the quadratic Casimir coincide. The adjoint representation is real for all $N\geq 2$.

{\bf Third consider the two index antisymmetric representation $2A$:} A $2A$ transformation is written as
\begin{align}
\tilde{q} \rightarrow \tilde{q}'  = U_{2A} \tilde{q} \ , \qquad  U_{2A} = 1+ i \alpha^a T^a_{2A}  + O\left(\alpha^2 \right)
\end{align}
or viewed as a tensor representation it is
\begin{align}
\tilde{q} \rightarrow \tilde{q}' = U_F \tilde{q} U_F^T \ , \qquad \tilde{q}^T = - \tilde{q} \ , \qquad U_F = 1+ i \alpha^a T_F^a + O \left( \alpha^2 \right)
\end{align}
The $2A$ representation is written as
\begin{align}
\left( \tilde{q} \right)^{ij} &= \tilde{q}^{r} \left( A^r \right)^{ij} \ , \qquad r=1,\ldots,\frac{N(N-1)}{2}
\end{align}
where we are expanding the two index antisymmetric representation in a set of linearly independent antisymmetric $N \times N$ matrices $\left( A^r \right)^T = - A^r$. There are $\frac{N(N-1)}{2}$ such matrices so $r=1,\ldots,\frac{N(N-1)}{2}$. We choose to normalize them similarly to the generators in the fundamental representation and so
\begin{align}
\text{Tr} \ A^r A^s  = T_F \delta^{rs} \ , \qquad \text{Tr} \ T_F^a T_F^b  = T_F \delta^{ab} 
\end{align}
This implies that the generators $T_{2A}^a$ are $ \frac{N(N-1)}{2} \times \frac{N(N-1)}{2}$ matrices and that the dimension of the two index antisymmetric representation is $\frac{N(N-1)}{2}$. The set of antisymmetric matrices $A^r$ satisfy the following completeness relation 
\begin{align}\label{eq:completenessA}
 \left( A^r \right)^{ij} \left( A^r \right)^{kl} &=  \frac{T_F}{2} \left( \delta^{il} \delta^{jk} - \delta^{ik} \delta^{jl} \right)
\end{align}
This will also be important to us so we will now prove it. Any complex antisymmetric $N \times N$ matrix $M$ can be written as a linear combination of $A^r$
\begin{align}
M = M^r A^r \ , \qquad r =1,\ldots,\frac{N(N-1)}{2}
\end{align}
for some complex coefficients $M^r$. Again we can project out these coefficients by multiplying with $A^s$ and taking the trace
\begin{align}
M^r &= \frac{1}{T_F} \text{Tr} \ MA^r
\end{align}
where we have used our conventions that the antisymmetric matrices are normalized as the generators in the fundamental representation. We can now write the expansion of the arbitrary antisymmetric matrix as
\begin{align}
M^{ij} &=  \frac{1}{T_F} M^{lk} \left( A^r \right)^{kl} \left( A^r \right)^{ij}
\end{align}
Writing the left-hand side as
\begin{align}
M^{ij} = \frac{1}{2} \left( \delta^{il} \delta^{jk} - \delta^{ik} \delta^{jl} \right) M^{lk}
\end{align}
we can put it together as
\begin{align}
\left[\frac{1}{T_F} \left( A^r\right)^{ij} \left( A^r\right)^{kl} - \frac{1}{2} \left( \delta^{il} \delta^{jk} - \delta^{ik} \delta^{jl} \right)  \right] M^{lk} = 0
\end{align}
Since $M$ is an arbitrary complex antisymmetric matrix this is only possible provided the square bracket vanishes and hence the identity above holds. For an infinitesimal $2A$ transformation we have
\begin{align}
\delta \tilde{q} = i \alpha^a \left(  T_F^a \tilde{q} + \tilde{q} \left( T_F^a \right)^T \right)  \ , \qquad  \delta \tilde{q}^{s} = i \alpha^a \left[ \frac{2}{T_F} \text{Tr}\ A^s T_F^a A^r \right] \tilde{q}^r
\end{align}
From this we can read off the generators
\begin{align}
\left( T_{2A}^a \right)^{sr} =  \frac{2}{T_F} \text{Tr} \ A^sT_F^a A^r 
\end{align}
Using the completeness relation Eq. \ref{eq:completenessA} we find the trace normalization factor
\begin{align}
\text{Tr} \  T^a_{2A} T_{2A}^b =T_{2A} \delta^{ab} \ , \qquad T_{2A} =  (2T_F) \frac{N-2}{2} 
\end{align}
As promised the trace form is diagonal and proportional to the identity and the trace normalization factor is fixed once $T_F$ is fixed. Finally we find the quadratic Casimir using completeness relations Eq. \ref{eq:completeness} and Eq. \ref{eq:completenessA}
\begin{align}
T_{2A}^a T_{2A}^a = C_{2A} \mathbb{I} \ , \qquad C_{2A} = (2T_F) \frac{(N+1)(N-2)}{N}
\end{align}
For $N=3$ the $2A$ representation is complex and equivalent to the complex conjugate fundamental representation $\bar{F}$, for $N=4$ it is strictly real $2A = \overline{2A}$ and for all $N \geq 5$ it is complex.

We now have the whole group theoretical machinery set up. It is convenient to choose $T_F = \frac{1}{2}$ for the fundamental representation. We here summarize our conventions
\begin{align}
\left( T_F \right)^{i}_{\phantom{i}j} &    &T_F& = \frac{1}{2}   &C_F& =  \frac{N^2-1}{2N}. &d_F& = N \\
\left( T_{\bar{F}} \right)_{i}^{\phantom{i}j} &= - \left( T_F^{a*} \right)_{i}^{\phantom{i}j}   &T_{\bar{F}}& = \frac{1}{2}   &C_{\bar{F}}& =  \frac{N^2-1}{2N}. &d_{\bar{F}}& = N \\
 \left(T_A^b \right)^{ac} &= i f^{abc}  &T_A& = N    &C_A& = N  &d_A& = N^2 -1 \\
 \left( T_{2A}^a \right)^{rs} &= 4 \text{Tr}\ A^r T_F^a A^s  &T_{2A}& = \frac{N-2}{2}   &C_{2A}& = \frac{(N+1)(N-2)}{N}  &d_{2A}& = \frac{N(N-1)}{2}
\end{align}
and the following two important identities 
\begin{align}
\label{eq:relations1}
 \left( T_F^a \right)^{i}_{\phantom{i}j} \left( T_F^a \right)^{k}_{\phantom{k}l} &=\frac{1}{2} \left(  \delta^{i}_{\phantom{i}l} \delta^{k}_{\phantom{k}j} - \frac{1}{N} \delta^{i}_{\phantom{i}j} \delta^{k}_{\phantom{k}l} \right) \\
 \left( A^r \right)^{ij} \left( A^r \right)^{kl} &=  \frac{1}{4 } \left( \delta^{il} \delta^{jk} - \delta^{ik} \delta^{jl} \right) \label{eq:relations2}
\end{align}

\section{Canonical Lagrangian}\label{app:canonicalL}

We now write down the canonically normalized Lagrangian of the theory in terms of the fields $A_{\mu}^a, q^i_{\alpha}, \tilde{q}^{r}_{\alpha}, Q_{\alpha,i,f}$
\begin{align}\label{eq:canonicalL}
\mathcal{L}  =& - \frac{1}{4} F_{\mu\nu}^a F^{a,\mu\nu}  + i \bar{q}_{\dot{\alpha},i} \bar{\sigma}^{\mu,\dot{\alpha}\alpha}  D^i_{\mu,j}  q_{\alpha}^{j} + i \bar{\tilde{q}}_{\dot{\alpha}}^r \bar{\sigma}^{\mu,\dot{\alpha}\alpha}  D^{rs}_{\mu}  \tilde{q}_{\alpha}^{s}  + i \bar{Q}^{i,f}_{\dot{\alpha}}   \bar{\sigma}^{\mu,\dot{\alpha}\alpha} D_{\mu,i}^{\phantom{\mu,i}j} Q_{\alpha,j,f}  \nonumber \\ 
& -\frac{1}{2\xi} \left( \partial^{\mu} A_{\mu}^a  \right)^2 + \bar{c}^a \left( -\partial^{\mu}  D_{\mu}^{ac} c^c \right)
\end{align}
where
\begin{align}
F_{\mu\nu}^a  &= \partial_{\mu} A_{\nu}^a - \partial_{\nu} A_{\mu}^a + g f^{abc} A_{\mu}^b A_{\nu}^c   \\
D^{i}_{\mu,j}  & = \partial_{\mu} \delta^{i}_{\phantom{i}j} -ig A_{\mu}^a \left( T^a_F \right)^{i}_{\phantom{i}j}. \\
D_{\mu}^{rs} & = \partial_{\mu} \delta^{rs} -ig A_{\mu}^a \left( T^a_{2A} \right)^{rs} \\
D_{\mu,i}^{\phantom{\mu,i}j} &=  \partial_{\mu} \delta_{i}^{\phantom{i}j} - i g A_{\mu}^a \left( T^a_{\bar{F}} \right)_{i}^{\phantom{i} j}  \\
D_{\mu}^{ac} & = \partial_{\mu} \delta^{ac} - i g A_{\mu}^b \left( T^b_A \right)^{ac}
\end{align}
The notation for Hermitian conjugation follows
\begin{align}
\left( A_{\mu}^{a}\right)^{\dagger}  = A_{\mu}^{a} \ , \qquad \left( q_{\alpha}^{i} \right)^{\dagger} \equiv \bar{q}_{\dot{\alpha},i} \ , \qquad \left( \tilde{q}_{\alpha}^r \right)^{\dagger} \equiv \bar{\tilde{q}}_{\dot{\alpha}}^{r}  \ , \qquad \left( Q_{\alpha,i,f} \right)^{\dagger} \equiv \bar{ Q}_{\dot{\alpha}}^{i,f} \ , \qquad \left( c^a \right)^{\dagger} = \bar{c}^a
\end{align}
We have also chosen linear covariant gauge with gauge parameter $\xi$ and $c^a$ are ghost fields in the adjoint representation. The propagators are the usual ones
\begin{align}\label{eq:canonicalP}
\left\langle A^a_{\mu}(x) A^b_{\nu}(y) \right\rangle & = \Delta_{\mu\nu} (x-y) \delta^{ab} \nonumber \\
\left\langle q^{i}_{\alpha}(x) \bar{q}_{\dot{\alpha},j} (y) \right\rangle &= S_{\alpha \dot{\alpha}}(x-y) \delta^{i}_{\phantom{i}j} \nonumber \\ 
\left\langle Q_{\alpha}^r(x) \bar{Q}_{\dot{\alpha}}^{s} (y) \right\rangle &= S_{\alpha \dot{\alpha}}(x-y) \delta^{rs} \\
\left\langle \tilde{q}_{\alpha,i,f}(x) \bar{Q}_{\dot{\alpha}}^{j,f'} (y) \right\rangle &= S_{\alpha \dot{\alpha}}(x-y) \delta_{i}^{\phantom{i}j}  \delta_{f}^{\phantom{f}f'} \nonumber \\
\left\langle c^{a}(x) \bar{c}^{b}(y) \right\rangle &= \Delta(x-y) \delta^{ab} \nonumber
\end{align}

We now switch to tensor notation for the gauge fields and $\tilde{q}$ Weyl fermions 
\begin{align}
A_{\mu} = A_{\mu}^a T_F^a \ ,\qquad  \tilde{q} = \tilde{q}^r A^r  \ ,\qquad c = c^a T^a_F
\end{align}
and rescale the fields appropriately
\begin{align}
A_{\mu} \rightarrow \frac{1}{g} A_{\mu} \ , \qquad q_{\alpha} \rightarrow \frac{1}{g} q_{\alpha} \ , \qquad \tilde{q}_{\alpha} \rightarrow \frac{1}{g} \tilde{q}_{\alpha} \ , \qquad Q_{\alpha} \rightarrow \frac{1}{g} Q_{\alpha} \ , \qquad c \rightarrow \frac{1}{g} c
\end{align}
With these two changes we can write the Lagrangian as
\begin{align}
\mathcal{L}  =& \frac{N_c}{\lambda} \left[ - \frac{1}{2} \text{Tr}\ F_{\mu\nu}F^{\mu\nu} + i \bar{q} \bar{\sigma}^{\mu} D_{\mu} q 
+ 2i \text{Tr} \ \left( \bar{\tilde{q}} \bar{\sigma}^{\mu} D_{\mu} \tilde{q} \right)  
+ i \bar{Q} \bar{\sigma}^{\mu} D_{\mu} Q  \right. \nonumber \\
& \left. - \frac{1}{\xi} \text{Tr} \ \left( \partial^{\mu} A_{\mu} \right)^2 
+ 2 \text{Tr} \ \bar{c} \left( - \partial^{\mu} D_{\mu} c \right)   \right]
\end{align}
with 
\begin{align}
F_{\mu\nu} &=  F_{\mu\nu}^{a} T^a_F= \partial_{\mu} A_{\nu} - \partial_{\nu} A_{\mu} - i \left[ A_{\mu},A_{\nu} \right] \\
D_{\mu} q^i & = \partial_{\mu} q^i -i \left( A_{\mu} \right)^{i}_{\phantom{i}j} q^j \ , \qquad \left( A_{\mu} \right)^{i}_{\phantom{i}j} = A_{\mu}^{a} \left( T^a_F \right)^{i}_{\phantom{i}j} \\
\left( D_{\mu} \tilde{q}  \right)^{ij}&= \partial_{\mu} \tilde{q}^{ij} -i \left( \left( A_{\mu} \right)^i_{\phantom{i}k} \tilde{q}^{kj} + \tilde{q}^{ik}\left( A_{\mu}^T \right)_k^{\phantom{k}j}   \right) \ , \qquad \left( A_{\mu} \right)^{i}_{\phantom{i}j} = A_{\mu}^{a} \left( T^a_F \right)^{i}_{\phantom{i}j} \\
D_{\mu} Q_i & = \partial_{\mu} Q_i -i \left( A_{\mu} \right)_{i}^{\phantom{i}j} Q_j \ , \qquad  \left( A_{\mu} \right)_{i}^{\phantom{i}j} = A_{\mu}^{a} \left( T^{a}_{\bar{F}} \right)_{i}^{\phantom{i}j} =  - A_{\mu}^{a} \left( T^{a,T}_{F} \right)_{i}^{\phantom{i}j}  \\
\left( D_{\mu} c \right)^{i}_{\phantom{i}j} &= \partial_{\mu} c^{i}_{\phantom{i}j} - i \left[ A_{\mu} , c\right]^{i}_{\phantom{i}j} \ , \qquad \left( A_{\mu} \right)^{i}_{\phantom{i}j} = A_{\mu}^{a} \left( T^a_F \right)^{i}_{\phantom{i}j} \ , \qquad \left( c \right)^{i}_{\phantom{i}j} = c^{a} \left( T^a_F \right)^{i}_{\phantom{i}j}
\end{align}
and the 't Hooft coupling being $\lambda = g^2 N_c$ and the gauge field is $A_{\mu} = A_{\mu}^a T^a_F$ for an appropriate representation $r$.

\newpage
\bibliography{Sample}

\end{document}